\documentclass[10pt,twocolumn,letterpaper]{article}
\usepackage{enumitem}  
\usepackage{cvpr}
\usepackage{times}
\usepackage{epsfig}
\usepackage{graphicx}
\usepackage{amsmath}
\usepackage{amssymb}
\usepackage{subcaption}
\usepackage{xcolor}
\usepackage{authblk}
\usepackage[title]{appendix}
\usepackage{multicol}
\usepackage{multirow}
\usepackage[pagebackref=true,breaklinks=true,letterpaper=true,colorlinks,bookmarks=false]{hyperref}

\newcommand{\tabincell}[2]{\begin{tabular}{@{}#1@{}}#2\end{tabular}} 

\cvprfinalcopy 

\def\cvprPaperID{3657} 

\ifcvprfinal\fi

\begin{document}

\title{DVC: An End-to-end Deep Video Compression Framework \vspace{-8mm}}

\author[1]{Guo Lu}
\author[2]{Wanli Ouyang}
\author[3]{Dong Xu}
\author[1]{Xiaoyun Zhang}
\author[1]{Chunlei  Cai}
\author[1]{Zhiyong Gao \thanks{Corresponding author}}
\affil[1]{Shanghai Jiao Tong University\\
       \{luguo2014, xiaoyun.zhang, caichunlei, zhiyong.gao\}@sjtu.edu.cn}
\affil[2]{The University of Sydney, SenseTime Computer Vision Research Group, Australia}       
\affil[3]{The University of Sydney\\
  \{wanli.ouyang, dong.xu\}@sydney.edu.au \vspace{-8mm}}

\maketitle
\thispagestyle{empty}

\begin{abstract}

Conventional video compression approaches use the predictive coding architecture and encode the corresponding motion information and residual information.
In this paper, taking advantage of both classical architecture in the conventional video compression method and the powerful non-linear representation ability of neural networks, we propose the first end-to-end video compression deep model that jointly optimizes all the components for video compression. 
Specifically, learning based optical flow estimation is utilized to obtain the motion information and reconstruct the current frames.
Then we employ two auto-encoder style neural networks to compress the corresponding motion and residual information.
All the modules are jointly learned through a single loss function, in which they collaborate with each other by considering the trade-off between reducing the number of compression bits and improving quality of the decoded video.
Experimental results show that the proposed approach can outperform the widely used video coding standard H.264 in terms of PSNR and be even on par with the latest standard H.265 in terms of MS-SSIM. Code is released at \href{https://github.com/GuoLusjtu/DVC}{https://github.com/GuoLusjtu/DVC}.

\end{abstract}

\vspace{-6mm}
\section{Introduction}

\begin{figure}[!t]
  \centering
  \begin{minipage}{0.45\linewidth}\footnotesize
    \centerline{\includegraphics[width=3.8cm]{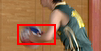}}
    \centerline{(a) Original frame (Bpp/MS-SSIM)}
  \end{minipage}
  \begin{minipage}{0.45\linewidth}\footnotesize
    \centerline{\includegraphics[width=3.8cm]{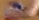}}
    \centerline{(b) H.264 (0.0540Bpp/0.945)}
  \end{minipage}
  
  \begin{minipage}{0.45\linewidth}\footnotesize
    \centerline{\includegraphics[width=3.8cm]{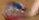}}
    \centerline{(c) H.265 (0.082Bpp/0.960)}
  \end{minipage}
  \begin{minipage}{0.45\linewidth}\footnotesize
    \centerline{\includegraphics[width=3.8cm]{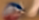}}
      \centerline{(d) Ours ( \textbf{0.0529Bpp}/ \textbf{0.961})}
  \end{minipage}
  
  \caption{Visual quality of the reconstructed frames from different video compression systems. (a) is the original frame. (b)-(d) are the reconstructed frames from H.264, H.265 and our method. Our proposed method only consumes 0.0529pp while achieving the best perceptual quality (0.961) when measured by MS-SSIM. (Best viewed in color.) }
\end{figure}

Nowadays, video content contributes to more than 80\% internet traffic \cite{networking2016forecast}, and the percentage is expected to increase even further. Therefore, it is critical to build an efficient video compression system and generate higher quality frames at given bandwidth budget.
In addition, most video related computer vision tasks such as video object detection or video object tracking are sensitive to the quality of compressed videos, and efficient video compression may bring benefits for other computer vision tasks.
Meanwhile, the techniques in video compression are also helpful for action recognition \cite{wu2018compressed} and model compression \cite{han2015deep}.

However, in the past decades, video compression algorithms \cite{wiegand2003overview,sullivan2012overview} rely on hand-crafted modules, e.g., block based motion estimation and Discrete Cosine Transform (DCT), to reduce the redundancies in the video sequences.
Although each module is well designed, the whole compression system is not end-to-end optimized. It is desirable to further improve video compression performance by jointly optimizing the whole compression system.

Recently, deep neural network (DNN) based auto-encoder for image compression \cite{toderici2015variable, balle2016end,toderici2017full, agustsson2017soft,balle2018variational,johnston2017improved,theis2017lossy,li2017learning,rippel2017real,agustsson2018generative} has achieved comparable or even better performance than the traditional image codecs like JPEG \cite{wallace1992jpeg}, JPEG2000 \cite{skodras2001jpeg} or BPG \cite{BPG}. 
One possible explanation is that the DNN based image compression methods can exploit large scale end-to-end training and highly non-linear transform, which are not used in the traditional approaches.

However, it is non-trivial to directly apply these techniques to build an end-to-end learning system for video compression.
\textbf{First}, it remains an open problem to learn how to generate and compress the motion information tailored for video compression. 
Video compression methods heavily rely on motion information to reduce temporal redundancy in video sequences.
A straightforward solution is to use the learning based optical flow to represent motion information.
However, current learning based optical flow approaches aim at generating flow fields as accurate as possible.
But, the precise optical flow is often not optimal for a particular video task \cite{xue2017video}.
In addition, the data volume of optical flow increases significantly when compared with motion information in the traditional compression systems and directly applying the existing compression approaches in \cite{wiegand2003overview, sullivan2012overview} to compress optical flow values will significantly increase the number of bits required for storing motion information.
\textbf{Second}, it is unclear how to build a DNN based video compression system by minimizing the rate-distortion based objective for both residual and motion information. Rate-distortion optimization (RDO) aims at achieving higher quality of reconstructed frame (\ie, less distortion) when the number of bits (or bit rate) for compression is given. RDO is important for video compression performance. In order to exploit the power of end-to-end training for learning based compression system, the RDO strategy is required to optimize the whole system.

In this paper, we propose the first end-to-end deep video compression (DVC) model that jointly learns motion estimation, motion compression, and residual compression. 
The advantages of the this network can be summarized as follows:
\begin{itemize}
    \setlength\itemsep{0em}
    \item All key components in video compression, \ie, motion estimation, motion compensation, residual compression,  motion compression,  quantization, and bit rate estimation, are implemented with an end-to-end neural networks. 
    \item The key components in video compression are jointly optimized based on rate-distortion trade-off through a single loss function, which leads to higher compression efficiency.    
    \item There is one-to-one mapping between the components of conventional video compression approaches and our proposed DVC model. This work serves as the bridge for researchers working on video compression, computer vision, and deep model design. For example, better model for optical flow estimation and image compression can be easily plugged into our framework.
    Researchers working on these fields can use our DVC model as a starting point for their future research.

\end{itemize}

Experimental results show that estimating and compressing motion information by using our neural network based approach can significantly improve the compression performance.
Our framework outperforms the widely used video codec H.264 when measured by PSNR and be on par with the latest video codec H.265 when measured by the multi-scale structural similarity index (MS-SSIM) \cite{wang2003multi}.

\section{Related Work}

\subsection{Image Compression}

A lot of image compression algorithms have been proposed in the past decades \cite{wallace1992jpeg, skodras2001jpeg, BPG}.
These methods heavily rely on handcrafted techniques.
For example, the JPEG standard linearly maps the pixels to another representation by using DCT, and quantizes the corresponding coefficients before entropy coding\cite{wallace1992jpeg}.
One disadvantage is that these modules are separately optimized and may not achieve optimal compression performance.

Recently, DNN based image compression methods have attracted more and more attention \cite{toderici2015variable,toderici2017full,balle2016end, balle2018variational, theis2017lossy, agustsson2017soft,li2017learning, rippel2017real,mentzer2018conditional,agustsson2018generative}.
In \cite{toderici2015variable,toderici2017full, johnston2017improved}, recurrent neural networks (RNNs) are utilized to build a progressive image compression scheme.
Other methods employed the CNNs to design an auto-encoder style network for image compression \cite{balle2016end, balle2018variational, theis2017lossy}.
To optimize the neural network, the work in \cite{toderici2015variable, toderici2017full, johnston2017improved} only tried to minimize the distortion (\eg, mean square error) between original frames and reconstructed frames without considering the number of bits used for compression.
Rate-distortion optimization technique was adopted in \cite{balle2016end, balle2018variational, theis2017lossy, li2017learning} for higher compression efficiency by introducing the number of bits in the optimization procedure.
To estimate the bit rates, context models are learned for the adaptive arithmetic coding method in \cite{rippel2017real,li2017learning,mentzer2018conditional}, while non-adaptive arithmetic coding is used in \cite{balle2016end, theis2017lossy}. 
In addition, other techniques such as generalized divisive normalization (GDN) \cite{balle2016end}, multi-scale image decomposition \cite{rippel2017real}, adversarial training \cite{rippel2017real}, importance map \cite{li2017learning, mentzer2018conditional} and intra prediction \cite{minnen2017spatially, baig2017learning} are proposed to improve the image compression performance. These existing works are important building blocks for our video compression network.

\subsection{Video Compression}

In the past decades, several traditional video compression algorithms have been proposed, such as H.264 \cite{wiegand2003overview} and H.265 \cite{sullivan2012overview}. Most of these algorithms follow the predictive coding architecture.
Although they provide highly efficient compression performance, they are manually designed and cannot be jointly optimized in an end-to-end way.

For the video compression task, a lot of DNN based methods have been proposed for intra prediction and residual coding\cite{chen2017deepcoder}, mode decision \cite{liu2016cu}, entropy coding \cite{song2017neural}, post-processing \cite{Lu_2018_ECCV}.  
These methods are used to improve the performance of one particular module of the traditional video compression algorithms instead of building an end-to-end compression scheme.
In \cite{chen2018learning}, Chen \textit{et al.} proposed a block based learning approach for video compression.
However, it will inevitably generate blockness artifact in the boundary between blocks.
In addition,  they used the motion information propagated from previous reconstructed frames through traditional block based motion estimation, which will degrade compression performance.
Tsai \textit{et al.} proposed an auto-encoder network to compress the residual from the H.264 encoder for specific domain videos \cite{tsai2018learning}.  This work does not use deep model for motion estimation, motion compensation or motion compression.

The most related work is the RNN based approach in  \cite{Wu_2018_ECCV}, where video compression is formulated as frame interpolation. 
However, the motion information in their approach is also generated by traditional block based motion estimation, which is encoded by the existing non-deep learning based image compression method \cite{WebP}. In other words,  estimation and compression of motion are not accomplished by deep model and jointly optimized with other components.
In addition, the video codec in \cite{Wu_2018_ECCV} only aims at minimizing the distortion (\ie, mean square error) between the original frame and reconstructed frame without considering rate-distortion trade-off in the training procedure.
In comparison, in our network, motion estimation and compression are achieved by DNN, which is jointly optimized with other components by considering the rate-distortion trade-off of the whole compression system.

\begin{figure*}[!t]
  \includegraphics[width=\linewidth]{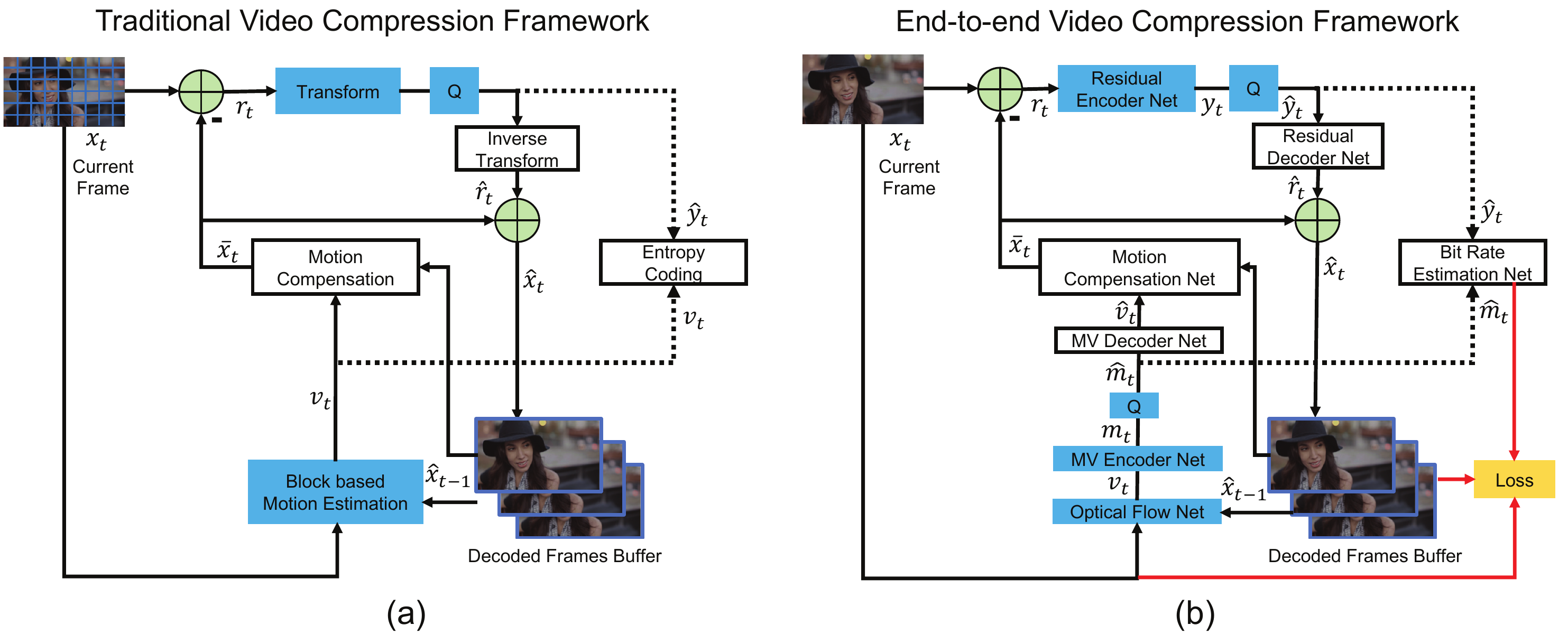}
  \caption{(a): The predictive coding architecture used by the traditional video codec H.264 \cite{wiegand2003overview} or H.265 \cite{sullivan2012overview}. (b): The proposed end-to-end video compression network. The modules with \textcolor{blue}{blue color} are not included in the decoder side.}
  \label{fig:overview}
\end{figure*}

\subsection{Motion Estimation}

Motion estimation is a key component in the video compression system. Traditional video codecs use the block based motion estimation algorithm \cite{wiegand2003overview}, which well supports hardware implementation.

In the computer vision tasks, optical flow is widely used to exploit temporal relationship.
Recently, a lot of learning based optical flow estimation methods \cite{dosovitskiy2015flownet, ranjan2017optical,sun2018pwc,hui2018liteflownet,hui2019lightweight} have been proposed. These approaches motivate us to integrate optical flow estimation into our end-to-end learning framework.
Compared with block based motion estimation method in the existing video compression approaches, learning based optical flow methods can provide accurate motion information at pixel-level, which can be also optimized in an end-to-end manner.
However, much more bits are required to compress motion information if optical flow values are encoded by traditional video compression approaches.

\section{Proposed Method}

\textbf{Introduction of Notations.}
Let $\mathcal{V}=\{x_1, x_2,..., x_{t-1}, x_t,...\}$ denote the current video sequences, where $x_t$ is the frame at time step $t$.
The predicted frame is denoted as $\bar{x}_t$ and the reconstructed/decoded frame is denoted as $\hat{x}_t$.
$r_t$ represents the residual (error) between the original frame $x_t$ and the predicted frame $\bar{x}_t$.
$\hat{r}_t$ represents the reconstructed/decoded residual.
In order to reduce temporal redundancy, motion information is required.
Among them, $v_t$ represents the motion vector or optical flow value. And $\hat{v}_t$ is its corresponding reconstructed version.
Linear or nonlinear transform can be used to improve the compression efficiency.
Therefore, residual information $r_t$ is transformed to $y_t$, and motion information $v_t$ can be transformed to $m_t$,
$\hat{r}_t$ and $\hat{m}_t$ are the corresponding quantized versions, respectively.

\subsection{Brief Introduction of Video Compression}
\label{sec:classic}
In this section, we give a brief introduction of video compression. More details are provided in \cite{wiegand2003overview,sullivan2012overview}.
Generally, the video compression encoder generates the bitstream based on the input current frames.
And the decoder reconstructs the video frames based on the received bitstreams.
In Fig. \ref{fig:overview}, all the modules are included in the encoder side while \textcolor{blue}{blue color} modules are not included in the decoder side.

The classic framework of video compression in Fig. \ref{fig:overview}(a) follows the predict-transform architecture.
Specifically, the input frame $x_{t}$ is split into a set of blocks, \ie, square regions, of the same size (\eg, $8\times 8$). 
The encoding procedure of the traditional video compression algorithm in the encoder side is shown as follows,

\textbf{Step 1. Motion estimation.} Estimate the motion between the current frame $x_t$ and the previous reconstructed frame $\Hat{x}_{t-1}$. The corresponding motion vector $v_t$ for each block is obtained.

\textbf{Step 2. Motion compensation.} The predicted frame $\bar{x}_t$ is obtained by copying the corresponding pixels in the previous reconstructed frame to the current frame based on the motion vector $v_t$ defined in Step 1. The residual $r_t$ between the original frame $x_t$ and the predicted frame $\bar{x}_t$ is obtained as $r_t = x_t - \bar{x}_t$. 

\textbf{Step 3. Transform and quantization.} The residual $r_t$ from Step 2 is quantized to $\hat{y}_t$. A linear transform (\eg, DCT) is used before quantization for better compression performance.

\textbf{Step 4. Inverse transform.}
The quantized result $\hat{y}_t$ in Step 3 is used by inverse transform for obtaining the reconstructed residual $\hat{r}_t$.

\textbf{Step 5. Entropy coding.} Both the motion vector $v_t$ in Step 1 and the quantized result $\hat{y}_t$ in Step 3 are encoded into bits by the entropy coding method and sent to the decoder.

\textbf{Step 6. Frame reconstruction.} The reconstructed frame $\Hat{x}_t$ is obtained by adding $\bar{x}_t$ in Step 2 and $\hat{r}_t$ in Step 4, \ie $\Hat{x}_t =\hat{r}_t + \bar{x}_t$. The reconstructed frame will be used by the $(t+1)$th frame at Step 1 for motion estimation.

For the decoder, based on the bits provided by the encoder at Step 5, motion compensation at Step 2, inverse quantization at Step 4, and then frame reconstruction at Step 6 are performed to obtain the reconstructed frame $\hat{x}_t$.

\subsection{Overview of the Proposed Method}

Fig. \ref{fig:overview} (b) provides a high-level overview of our end-to-end video compression framework. 
There is one-to-one correspondence between the traditional video compression framework and our proposed deep learning based framework.
The relationship and brief summarization on the differences are introduced as follows:

\textbf{Step N1. Motion estimation and compression.} We use a CNN model to estimate the optical flow \cite{ranjan2017optical}, which is considered as motion information ${v}_t$. Instead of directly encoding the raw optical flow values, an MV encoder-decoder network is proposed in Fig. \ref{fig:overview} to compress and decode the optical flow values, in which the quantized motion representation is denoted as $\hat{m}_t$. Then the corresponding reconstructed motion information $\hat{v}_t$ can be decoded by using the MV decoder net. Details are given in Section \ref{sec: mv_compress}.

\textbf{Step N2. Motion compensation.} A motion compensation network is designed to obtain the predicted frame $\bar{x}_t$ based on the optical flow obtained in Step N1. 
More information is provided in Section \ref{sec: mcnet}.

\textbf{Step N3-N4. Transform, quantization and inverse transform.}
We replace the linear transform in Step 3 by using a highly non-linear residual encoder-decoder network, and the residual $r_t$ is non-linearly maped to the representation $y_t$.
Then $y_t$ is quantized to $\hat{y}_t$.
In order to build an end-to-end training scheme, we use the quantization method in \cite{balle2016end}. 
The quantized representation $\hat{y}_t$ is fed into the residual decoder network to obtain the reconstructed residual $\hat{r}_t$.
Details are presented in Section \ref{sec:res_codec} and \ref{sec:training}.

\textbf{Step N5. Entropy coding.}
At the testing stage, the quantized motion representation $\hat{m}_t$ from Step N1 and the residual representation $\hat{y}_t$ from Step N3 are coded into bits and sent to the decoder.
At the training stage, to estimate the number of bits cost in our proposed approach, we use the CNNs (Bit rate estimation net in Fig. \ref{fig:overview}) to obtain the probability distribution of each symbol in $\hat{m}_t$ and $\hat{y}_t$.
More information is provided in Section \ref{sec:training}.

\textbf{Step N6. Frame reconstruction.} It is the same as Step 6 in Section \ref{sec:classic}.

\subsection{MV Encoder and Decoder Network}
\label{sec: mv_compress}

\begin{figure}[!t]
  \includegraphics[width=\linewidth]{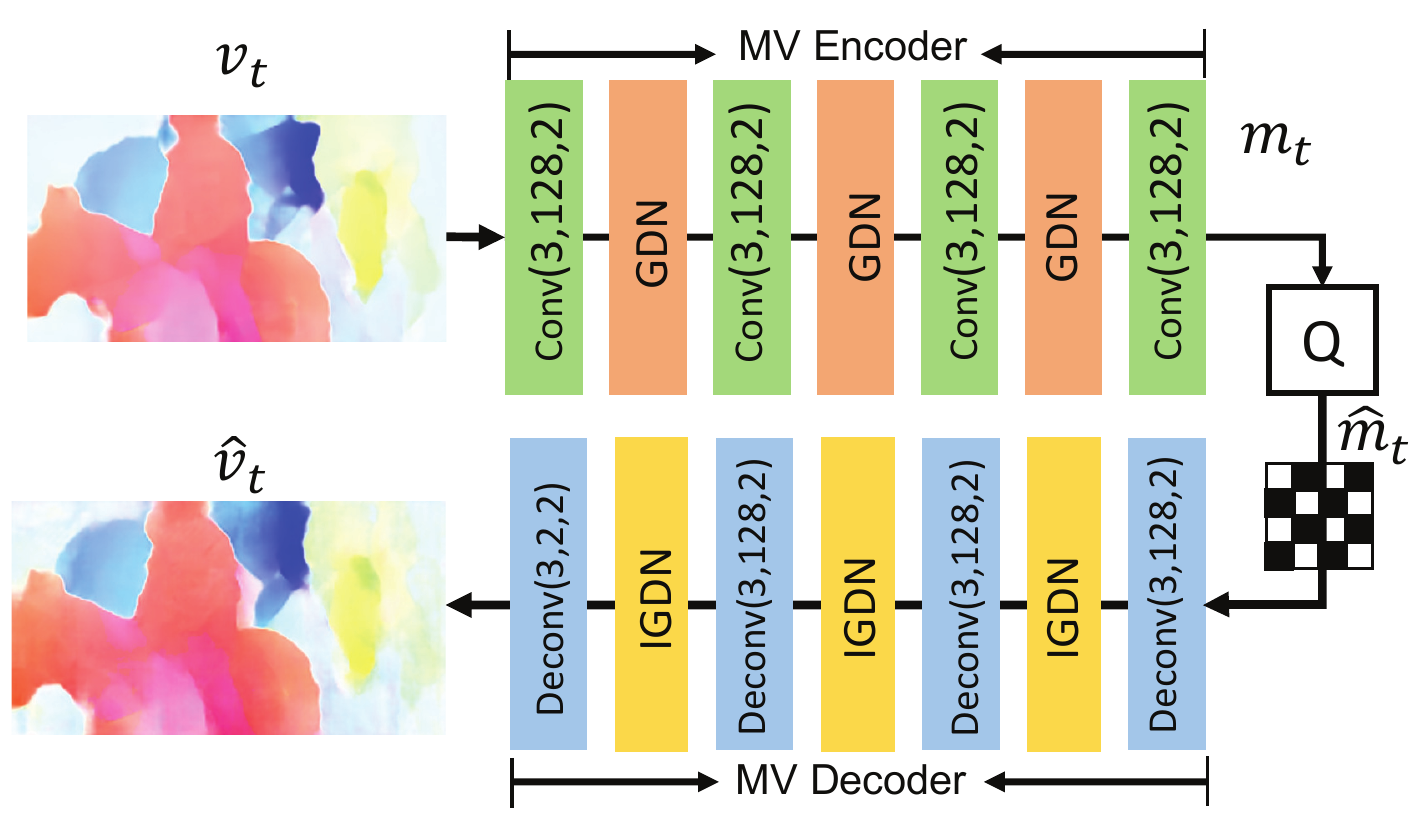}
  \caption{Our MV Encoder-decoder network. Conv(3,128,2) represents the convoluation operation with the kernel size of 3x3, the output channel of 128 and the stride of 2. GDN/IGDN \cite{balle2016end} is the nonlinear transform function. The binary feature map is only used for illustration. }
  \label{fig:mvencoder}
\end{figure}

In order to compress motion information at Step N1, we design a CNN to transform the optical flow to the corresponding representations for better encoding.
Specifically, we utilize an auto-encoder style network to compress the optical flow, which is first proposed by \cite{balle2016end} for the image compression task.
The whole MV compression network is shown in Fig. \ref{fig:mvencoder}. 
The optical flow $v_t$ is fed into a series of convolution operation and nonlinear transform. The number of output channels for convolution (deconvolution) is 128 except for the last deconvolution layer, which is equal to 2.
Given optical flow $v_{t}$ with the size of $M \times N \times 2$, the MV encoder will generate the motion representation $m_{t}$ with the size of $M/16 \times N/16 \times 128$. Then $m_{t}$ is quantized to $\hat{m}_{t}$. 
The MV decoder receives the quantized representation and reconstruct motion information $\hat{v}_t$.
In addition, the quantized representation $ \hat{m}_t$ will be used for entropy coding.

\subsection{Motion Compensation Network}
\label{sec: mcnet}
Given the previous reconstructed frame $\hat{x}_{t-1}$ and the motion vector $\hat{v}_t$, the motion compensation network obtains the predicted frame $\bar{x}_t$, which is expected to as close to the current frame $x_t$ as possible.  
First, the previous frame $\hat{x}_{t-1}$ is warped to the current frame based on the motion information $\hat{v}_{t}$. The warped frame still has artifacts. To remove the artifacts, we concatenate the warped frame $w(\hat{x}_{t-1}, \hat{v}_{t})$, the reference frame $\hat{x}_{t-1}$, and the motion vector $\hat{v}_t$ as the input, then feed them into another CNN to obtain the refined predicted frame $\bar{x}_t$.
The overall architecture of the proposed network is shown in Fig. \ref{fig:mc_net}.
The detail of the CNN in Fig. \ref{fig:mc_net} is provided in supplementary material.
Our proposed method is a pixel-wise motion compensation approach, which can provide more accurate temporal information and avoid the blockness artifact in the traditional block based motion compensation method.
It means that we do not need the hand-crafted loop filter or the sample adaptive offset technique \cite{wiegand2003overview, sullivan2012overview} for post processing.

\subsection{Residual Encoder and Decoder Network}
\label{sec:res_codec}
The residual information $r_t$ between the original frame $x_t$ and the predicted frame $\bar{x}_t$ is encoded by the residual encoder network as shown in Fig. \ref{fig:overview}.
In this paper, we rely on the highly non-linear neural network in \cite{balle2018variational} to transform the residuals to the corresponding latent representation.
Compared with discrete cosine transform in the traditional video compression system, our approach can better exploit the power of non-linear transform and achieve higher compression efficiency.

\subsection{Training Strategy}
\label{sec:training}

\textbf{Loss Function.}
The goal of our video compression framework is to minimize the number of bits used for encoding the video, while at the same time reduce the distortion between the original input frame $x_t$ and the reconstructed frame $\hat{x}_t$. Therefore, we propose the following rate-distortion optimization problem,
\begin{equation}
    \lambda D+ R = \lambda d(x_t, \hat{x}_t) +  (H(\hat{m}_t) + H(\hat{y}_t)),
\label{eq:rdo}    
\end{equation}
where $d(x_t, \hat{x}_t)$ denotes the distortion between $x_t$ and $\hat{x}_t$, and we use mean square error (MSE) in our implementation.
$H(\cdot)$ represents the number of bits used for encoding the representations. In our approach, both residual representation $\hat{y}_t$ and motion representation $\hat{m}_t$ should be encoded into the bitstreams. $\lambda$ is the Lagrange multiplier that determines the trade-off between the number of bits and distortion.
As shown in Fig. \ref{fig:overview}(b), the reconstructed frame $\hat{x}_t$, the original frame $x_{t}$ and the estimated bits are input to the loss function.

\begin{figure}[!t]
  \includegraphics[width=\linewidth]{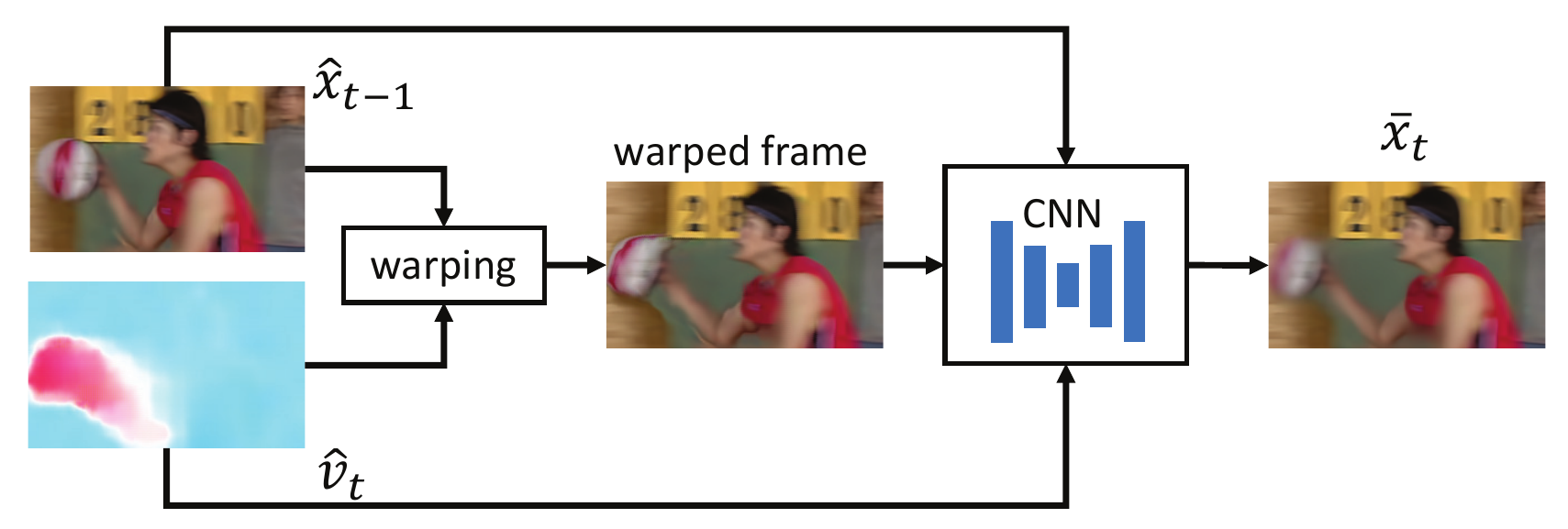}
  \caption{Our Motion Compensation Network.}
  \label{fig:mc_net}
\end{figure}

\begin{figure*}[!t]
  \centering
  \begin{minipage}{0.30\textwidth}
    \centerline{\includegraphics[width=5cm]{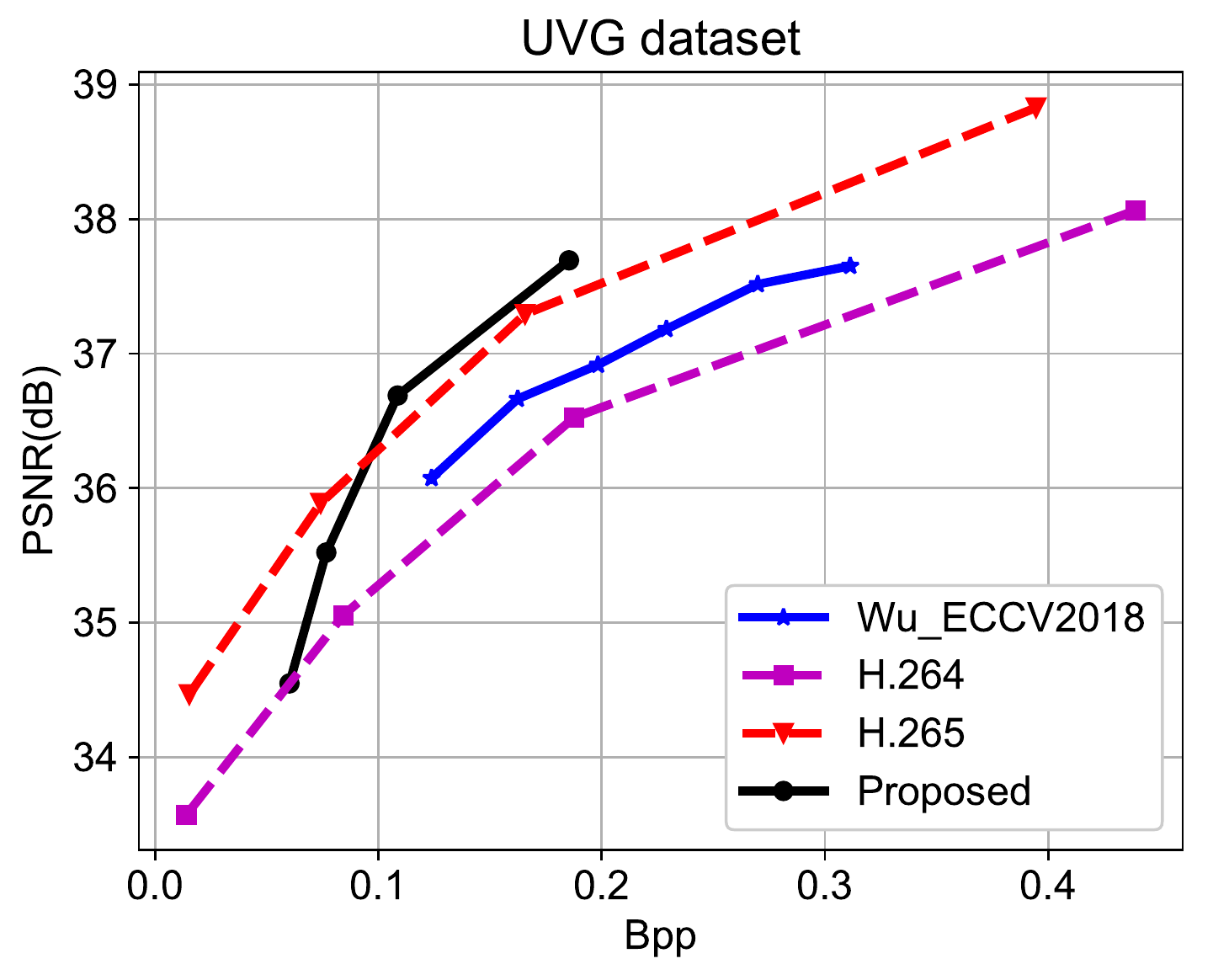}}
  \end{minipage} 
  \begin{minipage}{0.30\textwidth}
    \centerline{\includegraphics[width=5cm]{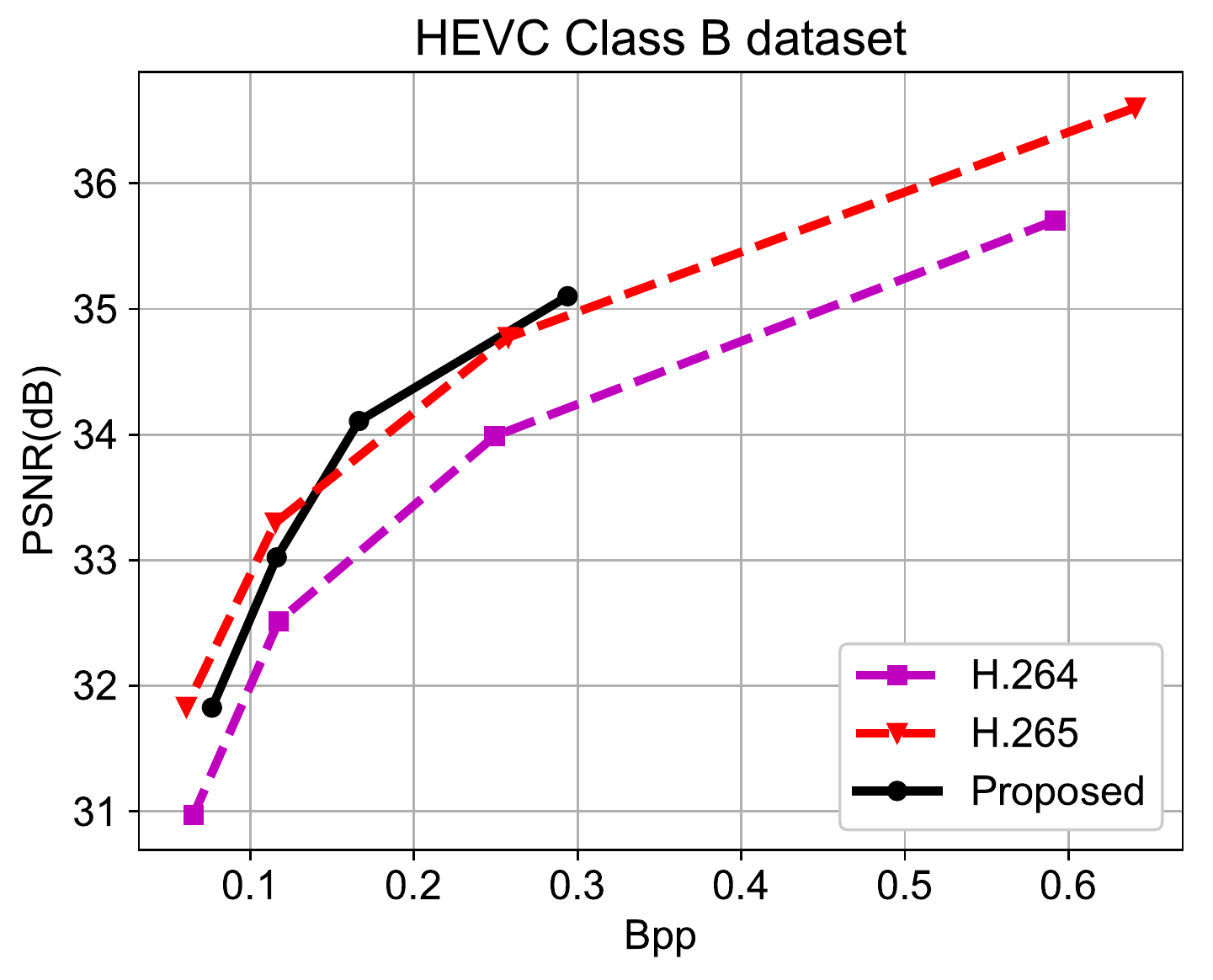}}
  \end{minipage}
    \begin{minipage}{0.30\textwidth}
    \centerline{\includegraphics[width=5cm]{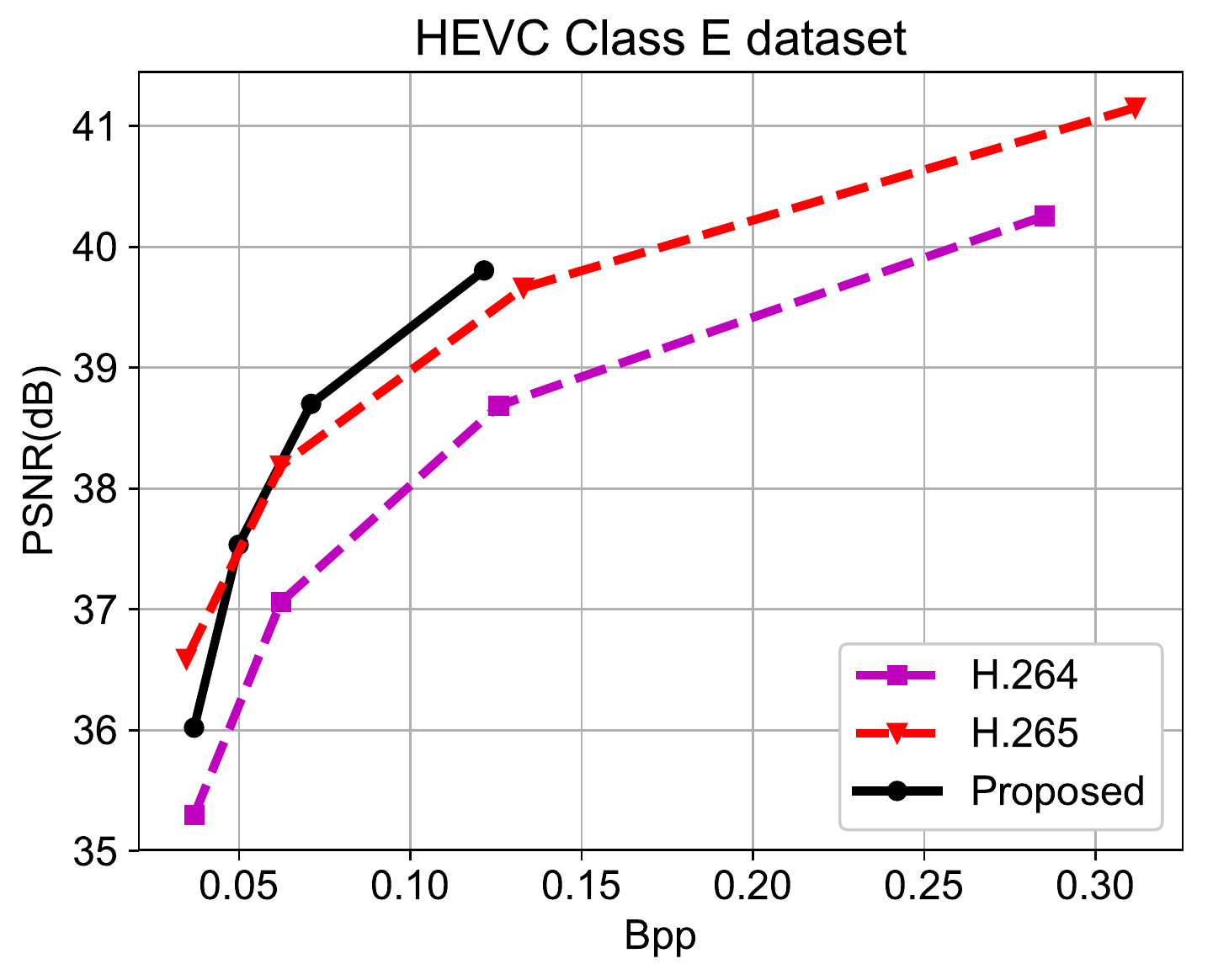}}
  \end{minipage}

    \begin{minipage}{0.30\textwidth}
    \centerline{\includegraphics[width=5cm]{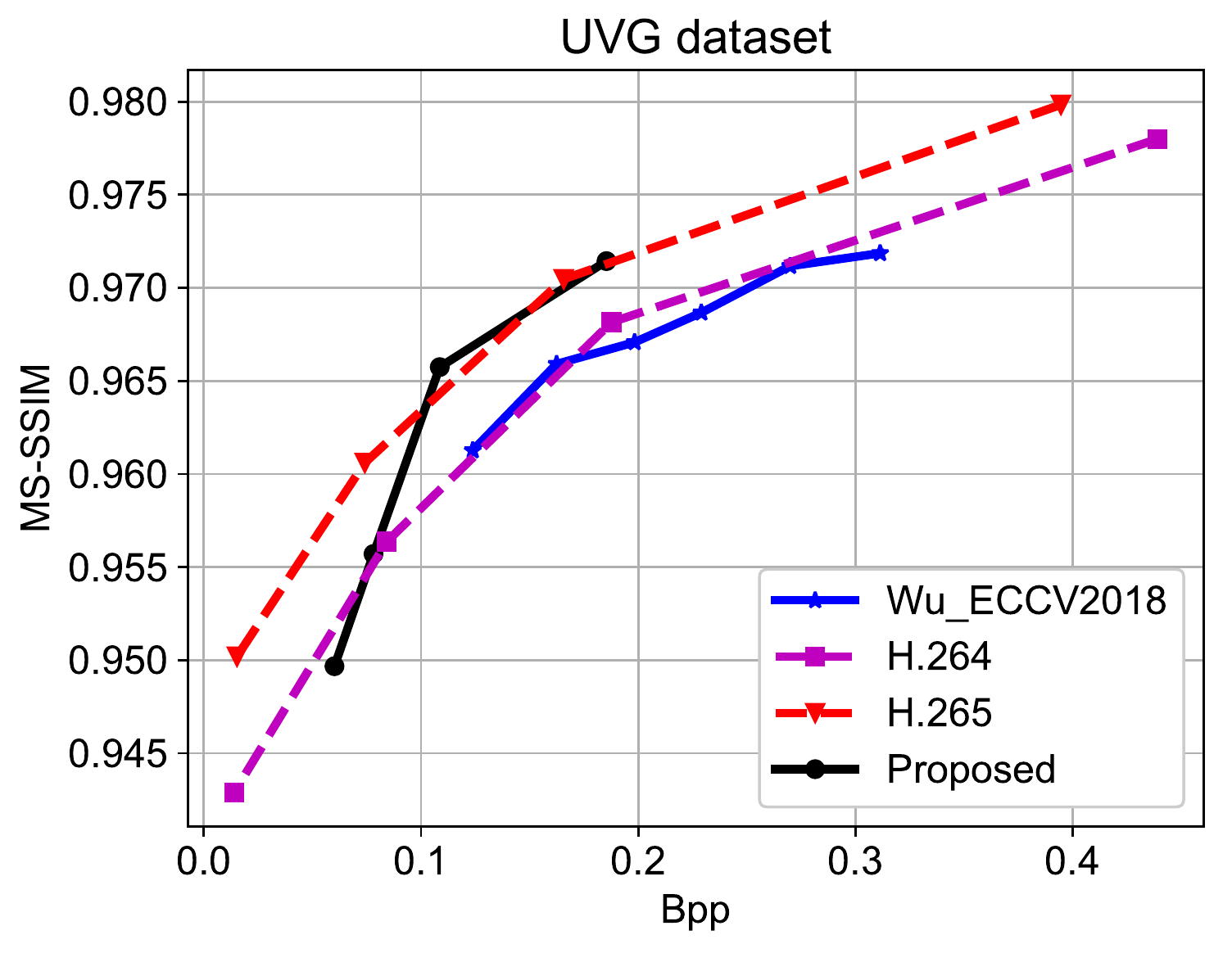}}
  \end{minipage}
  \begin{minipage}{0.30\textwidth}
    \centerline{\includegraphics[width=5cm]{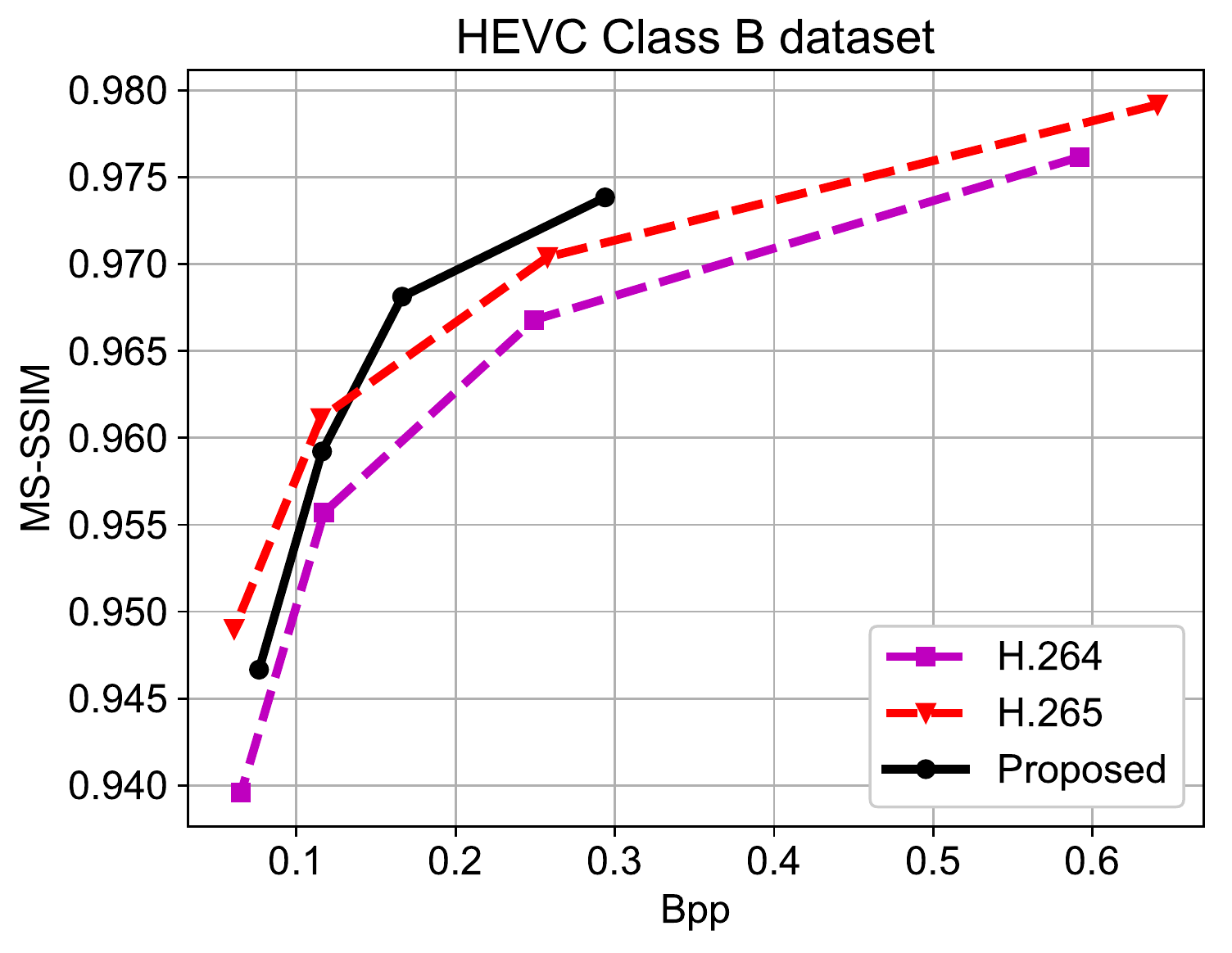}}
  \end{minipage}
    \begin{minipage}{0.30\textwidth}
    \centerline{\includegraphics[width=5cm]{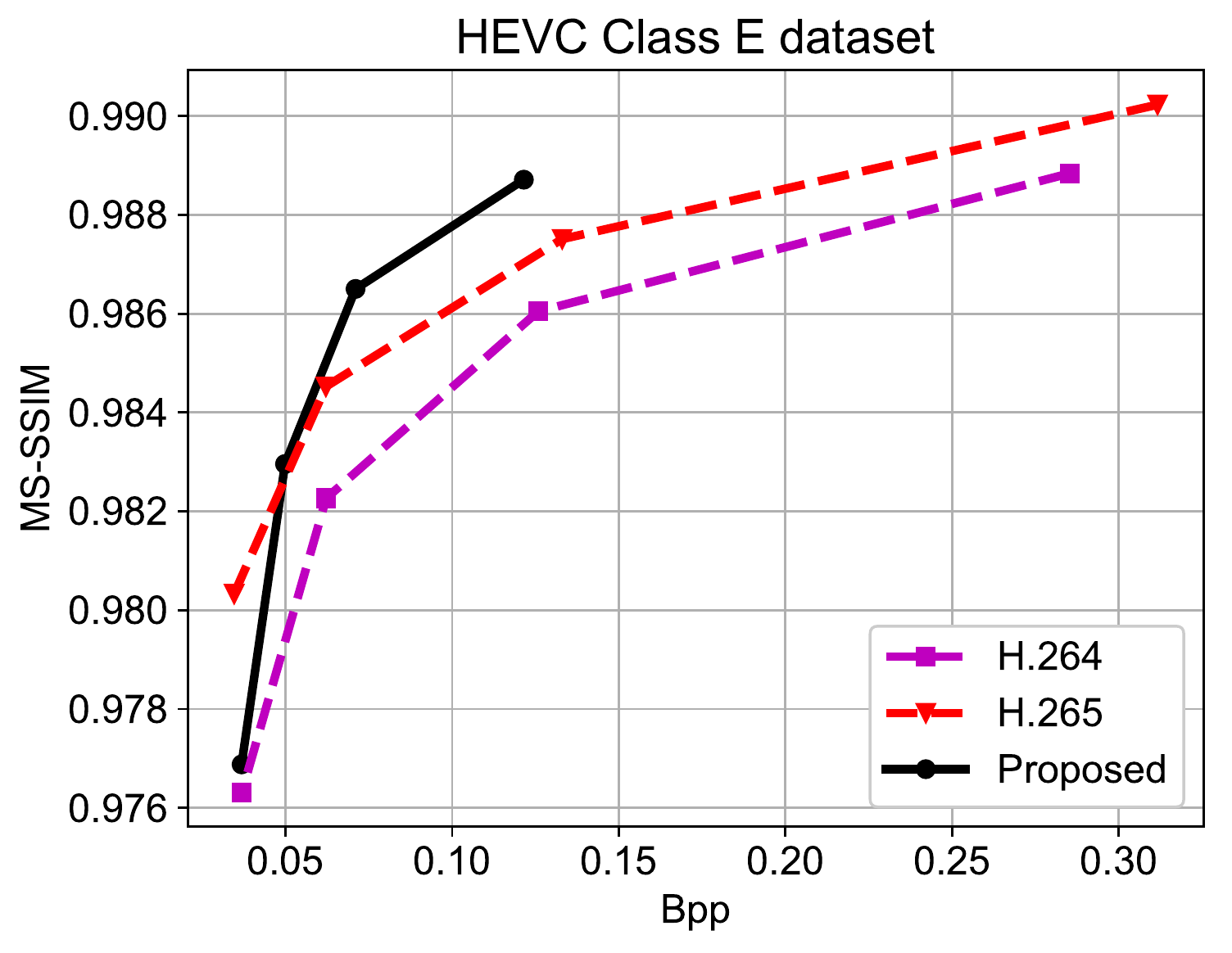}}
  \end{minipage}
  
  \caption{Comparsion between our proposed method with the learning based video codec in \cite{Wu_2018_ECCV}, H.264 \cite{wiegand2003overview} and H.265 \cite{sullivan2012overview}. Our method outperforms H.264 when measured by both PSNR ans MS-SSIM. Meanwhile, our method achieves similar or better compression performance when compared with H.265 in terms of MS-SSIM.}
  
  \label{fig:mainresults}
\end{figure*}

\textbf{Quantization.}
Latent representations such as residual representation $y_t$ and motion representation $m_t$ are required to be quantized before entropy coding. However, quantization operation is not differential, which makes end-to-end training impossible.
To address this problem, a lot of methods have been proposed \cite{toderici2015variable, agustsson2017soft, balle2016end}. 
In this paper, we use the method in \cite{balle2016end} and replace the quantization operation by adding uniform noise in the training stage.
Take $y_t$ as an example, the quantized representation $\hat{y}_t$ in the training stage is approximated by adding uniform noise to $y_t$, \ie, $\hat{y}_t=y_t + \eta$, where $\eta$ is uniform noise.
In the inference stage, we use the rounding operation directly, \ie, $\hat{y}_t= round(y_t)$.

\textbf{Bit Rate Estimation.}
In order to optimize the whole network for both number of bits and distortion, we need to obtain the bit rate ($H(\hat{y}_t)$ and $H(\hat{m}_t)$) of the generated latent representations ($\hat{y}_t$ and $\hat{m}_t$).
The correct measure for bitrate is the entropy of the corresponding latent representation symbols.
Therefore, we can estimate the probability distributions of $\hat{y}_t$ and $\hat{m}_t$, and then obtain the corresponding entropy. 
In this paper, we use the CNNs in \cite{balle2018variational} to estimate the distributions. 

\textbf{Buffering Previous Frames.}
As shown in Fig. \ref{fig:overview}, the previous reconstructed frame $\hat{x}_{t-1}$ is required in the motion estimation and motion compensation network when compressing the current frame. However, the previous reconstructed frame $\hat{x}_{t-1}$ is the output of our network for the previous frame, which is based on the reconstructed frame $\hat{x}_{t-2}$, and so on.  Therefore, the frames $x_1, \ldots, x_{t-1}$ might be required during the training procedure for the frame $x_t$, which reduces the variation of training samples in a mini-batch and could be impossible to be stored in a GPU when $t$ is large.
To solve this problem, we adopt an on line updating strategy.
Specifically, the reconstructed frame $\hat{x}_t$ in each iteration will be saved in a buffer. In the following iterations, $\hat{x}_{t}$ in the buffer will be used for motion estimation and motion compensation when encoding $x_{t+1}$.
Therefore, each training sample in the buffer will be updated in an epoch. In this way, we can optimize and store one frame for a video clip in each iteration, which is more efficient.

\section{Experiments}

\subsection{Experimental Setup}
\textbf{Datasets.} We train the proposed video compression framework using the Vimeo-90k dataset \cite{xue2017video}, which is recently built for evaluating different video processing tasks, such as video denoising and video super-resolution.
It consists of 89,800 independent clips that are different from each other in content.

To report the performance of our proposed method, we evaluate our proposed algorithm on the UVG dataset \cite{UVG}, and the HEVC Standard Test Sequences (Class B, Class C, Class D and Class E) \cite{sullivan2012overview}.
The content and resolution of these datasets are diversified and they are widely used to measure the performance of video compression algorithms. 

\textbf{Evaluation Method}   
To measure the distortion of the reconstructed frames, we use two evaluation metrics: PSNR and MS-SSIM \cite{wang2003multi}. 
MS-SSIM correlates better with human perception of distortion than PSNR.
To measure the number of bits for encoding the representations, we use bits per pixel(Bpp) to represent the required bits for each pixel in the current frame.

\textbf{Implementation Details}
We train four models with different $\lambda$ ($\lambda = 256, 512, 1024, 2048$).
For each model, we use the Adam optimizer \cite{kingma2014adam} by setting the initial learning rate as 0.0001, $\beta_1$ as 0.9 and $\beta_2$ as 0.999, respectively.
The learning rate is divided by 10 when the loss becomes stable. 
The mini-batch size is set as 4. 
The resolution of training images is $256 \times 256$. 
The motion estimation module is initialized with the pretrained weights in \cite{ranjan2017optical}.
The whole system is implemented based on Tensorflow and it takes about 7 days to train the whole network using two Titan X GPUs.

\subsection{Experimental Results }
In this section, both H.264 \cite{wiegand2003overview} and H.265 \cite{sullivan2012overview} are included for comparison. In addition, a learning based video compression system in \cite{Wu_2018_ECCV}, denoted by Wu\_ECCV2018, is also included for comparison.
To generate the compressed frames by the H.264 and H.265, we follow the setting in \cite{Wu_2018_ECCV} and use the FFmpeg with \textit{very fast} mode. The GOP sizes for the UVG dataset and HEVC dataset are 12 and 10, respectively.
Please refer to supplementary material for more details about the H.264/H.265 settings.

Fig. \ref{fig:mainresults} shows the experimental results on the UVG dataset, the HEVC Class B and Class E datasets.
The results for HEVC Class C and Class D are provided in supplementary material.
It is obvious that our method outperforms the recent work on video compression \cite{Wu_2018_ECCV} by a large margin.
On the UVG dataset, the proposed method achieved about 0.6dB gain at the same Bpp level.
It should be mentioned that our method only uses one previous reference frame while the work by Wu \textit{et al.} \cite{Wu_2018_ECCV} utilizes bidirectional frame prediction and requires two neighbouring frames.
Therefore, it is possible to further improve the compression performance of our framework by exploiting temporal information from multiple reference frames.

On most of the datasets, our proposed framework outperforms the H.264 standard when measured by PSNR and MS-SSIM.
In addition, our method achieves similar or better compression performance when compared with H.265 in terms of MS-SSIM.
As mentioned before, the distortion term in our loss function is measured by MSE. Nevertheless, our method can still provide reasonable visual quality in terms of MS-SSIM.

\begin{figure}[!t]
  \centering
  \begin{minipage}{0.45\linewidth}\footnotesize
    \centerline{\includegraphics[width=8.2cm]{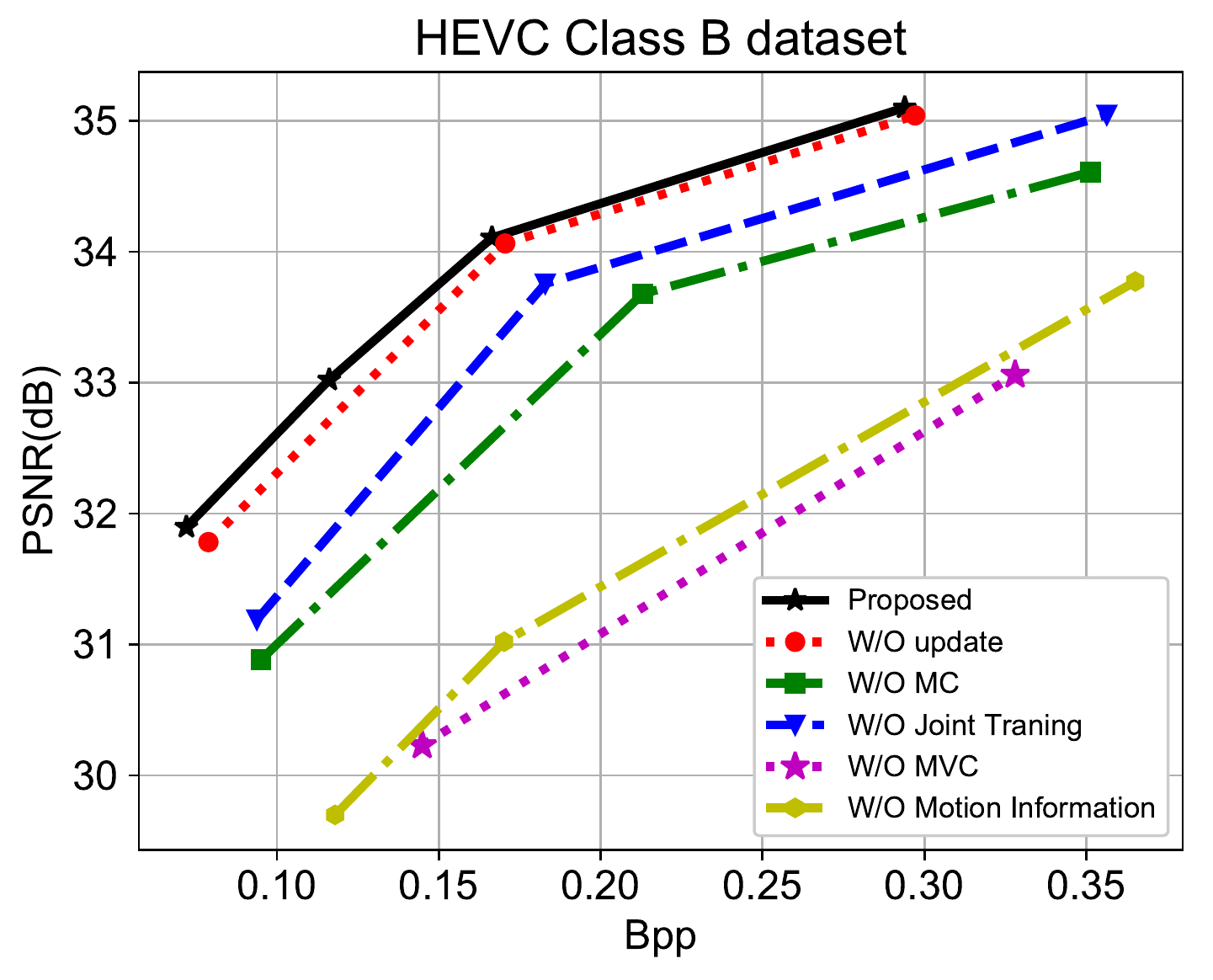}}
  \end{minipage}

  \caption{Ablation study. We report the compression performance in the following settings. 1. The strategy of buffering previous frame is not adopted(\textcolor{red}{W/O update}). 2. Motion compensation network is removed (\textcolor{green}{W/O MC}). 3. Motion estimation module is not jointly optimized (\textcolor{blue}{W/O Joint Training}). 4. Motion compression network is removed (\textcolor{magenta}{W/O MVC}). 5. Without relying on motion information (\textcolor[rgb]{0.749,0.749,0.239}{W/O Motion Information}).}
  \label{fig:ablation}
\end{figure}

\subsection{Ablation Study and Model Analysis}
\label{sec:ablation}
\textbf{Motion Estimation.}
In our proposed method, we exploit the advantage of the end-to-end training strategy and optimize the motion estimation module within the whole network. 
Therefore, based on rate-distortion optimization, the optical flow in our system is expected to be more compressible, leading to more accurate warped frames.
To demonstrate the effectiveness, we perform a experiment by fixing the parameters of the initialized motion estimation module in the whole training stage. In this case, the motion estimation module is pretrained only for estimating optical flow more accurately, but not for optimal rate-distortion.
Experimental result in Fig. \ref{fig:ablation} shows that our approach with joint training for motion estimation can improve the performance significantly when compared with the approach by fixing motion estimation, which is the denoted by \emph{W/O Joint Training} in Fig. \ref{fig:ablation} (see the \textcolor{blue}{blue curve}).

We report the average bits costs for encoding the optical flow and the corresponding PSNR of the warped frame in Table \ref{tab:jointMV}. 
Specifically,
when the motion estimation module is fixed during the training stage, it needs 0.044bpp to encode the generated optical flow and the corresponding PSNR of the warped frame is 27.33db. 
In contrast, we need 0.029bpp to encode the optical flow in our proposed method, and the PSNR of warped frame is higher (28.17dB). Therefore, the joint learning strategy not only saves the number of bits required for encoding the motion, but also has better warped image quality. These experimental results clearly show that putting motion estimation into the rate-distortion optimization improves compression performance.

\begin{figure}[!t]
  \captionsetup[subfigure]{aboveskip=-0.8pt,belowskip=-1pt}
  \centering
  \begin{subfigure}[t]{0.235\textwidth}\footnotesize
  \centering
   \includegraphics[width=\linewidth]{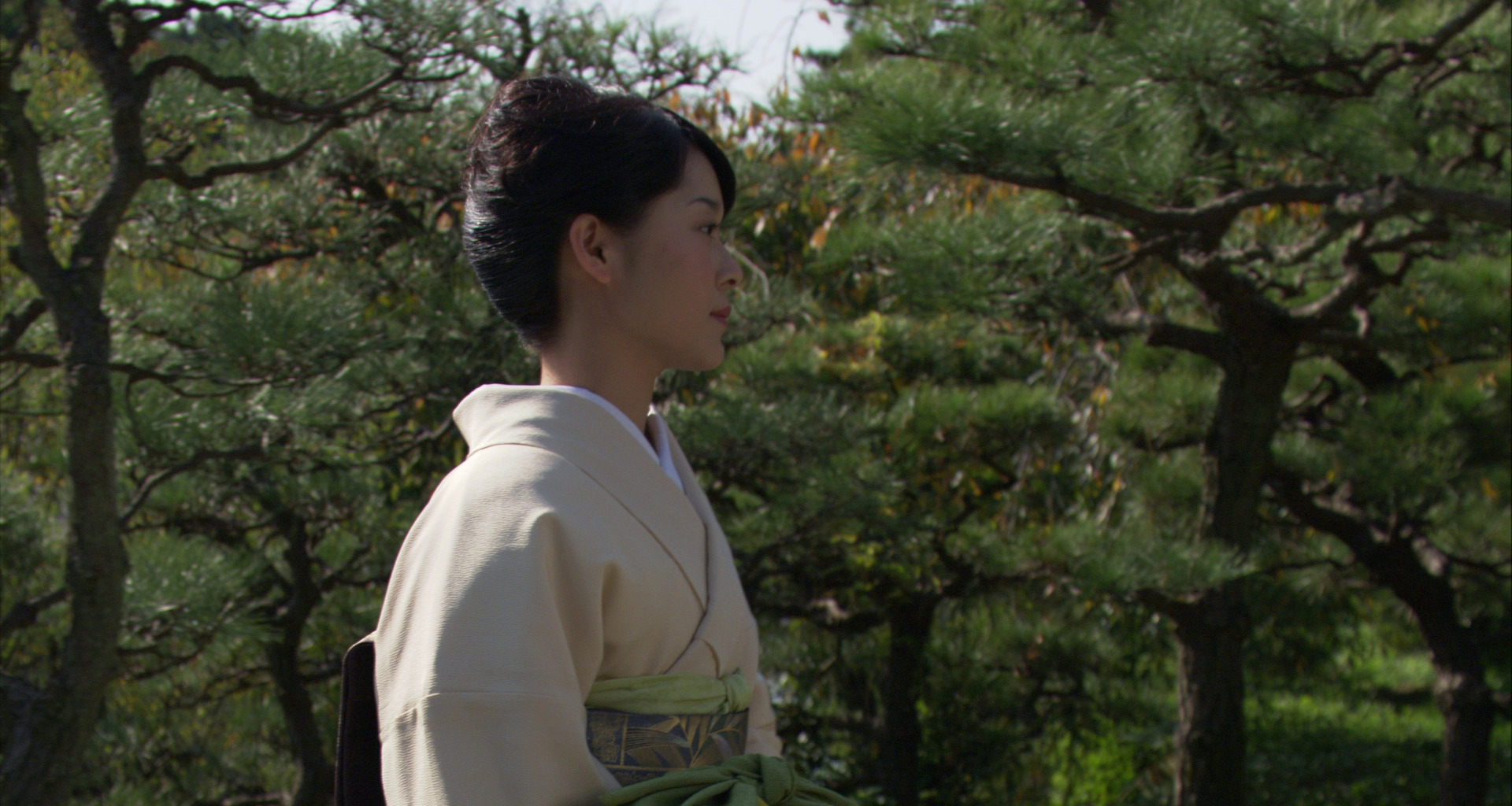}
      \caption{Frame No.5} 
  \end{subfigure}
\hfill    
  \begin{subfigure}[t]{0.235\textwidth}\footnotesize
  \centering
   \includegraphics[width=\linewidth]{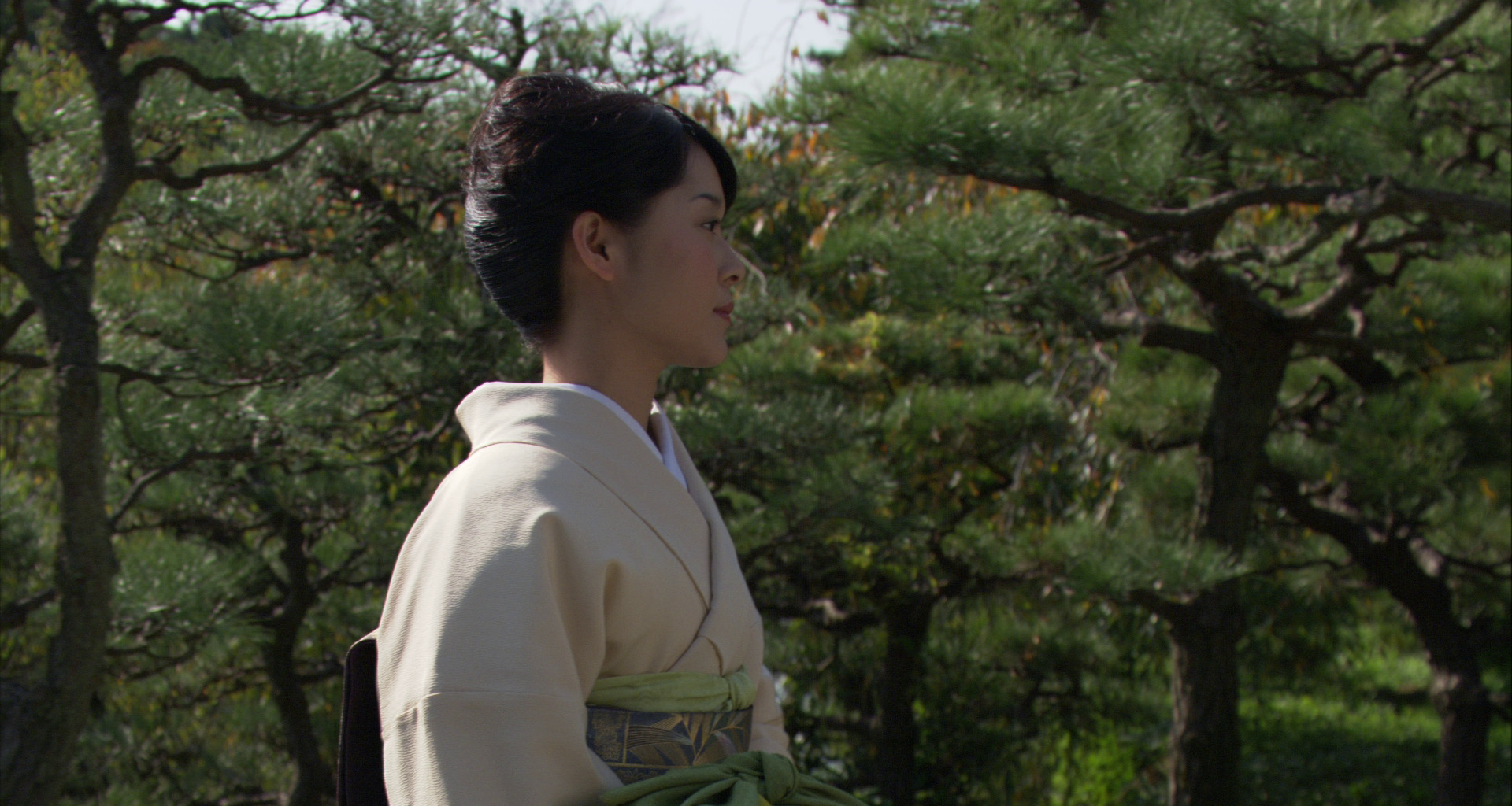}
      \caption{Frame No.6} 
  \end{subfigure}

  \begin{subfigure}[t]{0.235\textwidth}\footnotesize
  
  \centering
   \includegraphics[width=\linewidth]{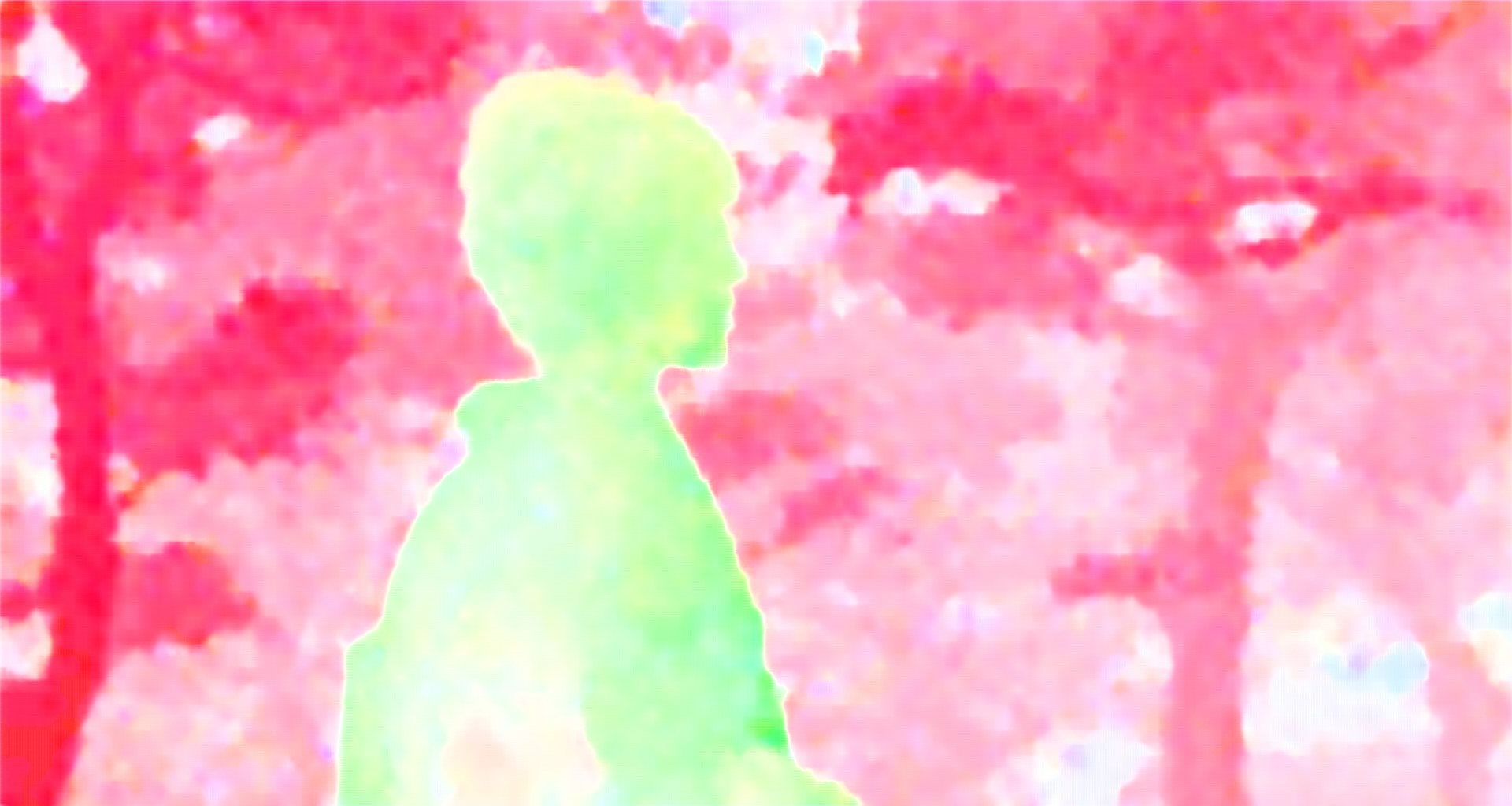}
      \caption{Reconstructed optical flow when fixing ME Net.} 
  \end{subfigure}
\hfill    
  \begin{subfigure}[t]{0.235\textwidth}\footnotesize
  \centering
   \includegraphics[width=\linewidth]{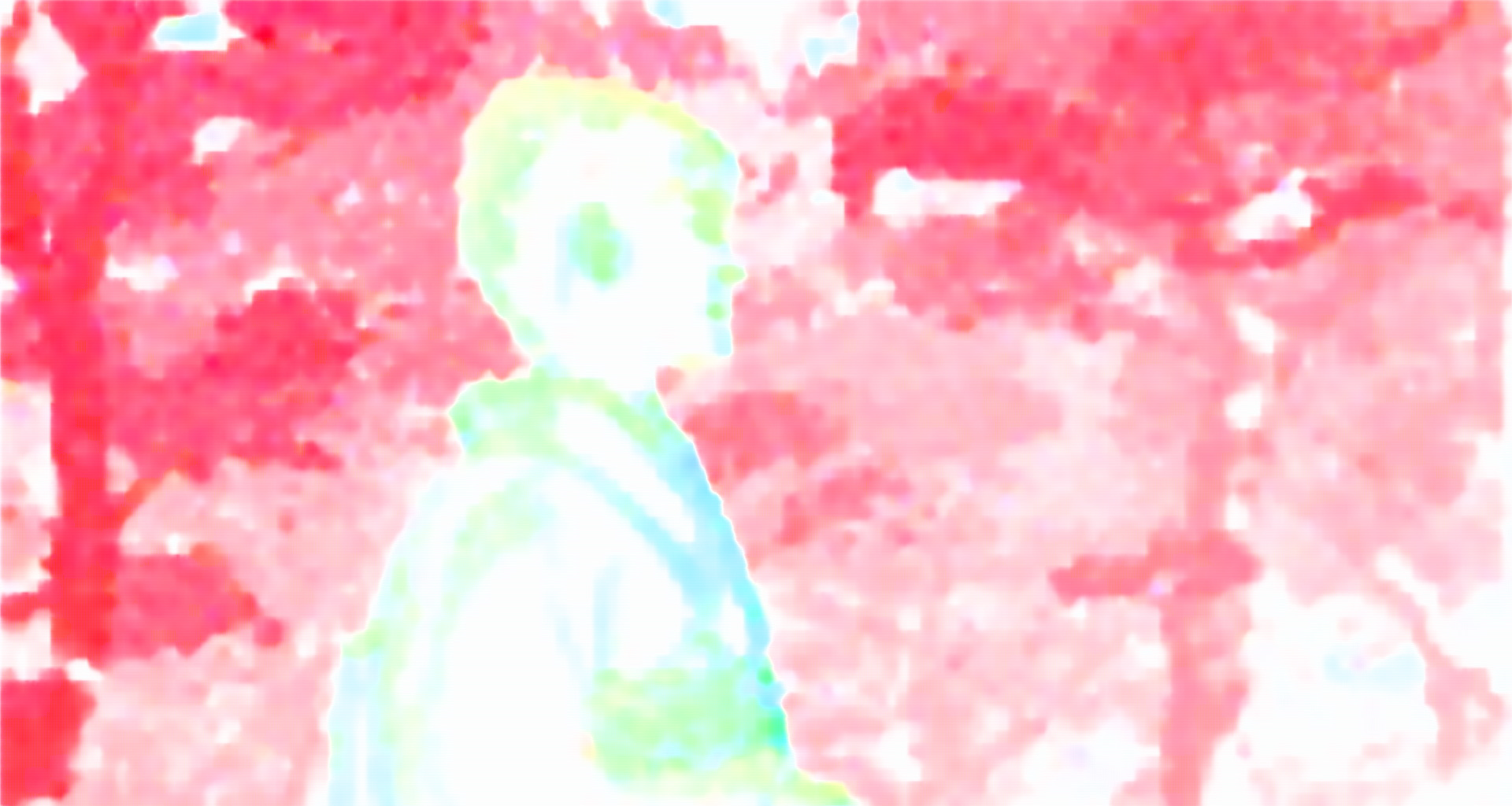}
      \caption{Reconstructed optical flow with the joint training strategy.} 
  \end{subfigure}

  \begin{subfigure}[t]{0.235\textwidth}\footnotesize
  \centering
   \includegraphics[width=\linewidth]{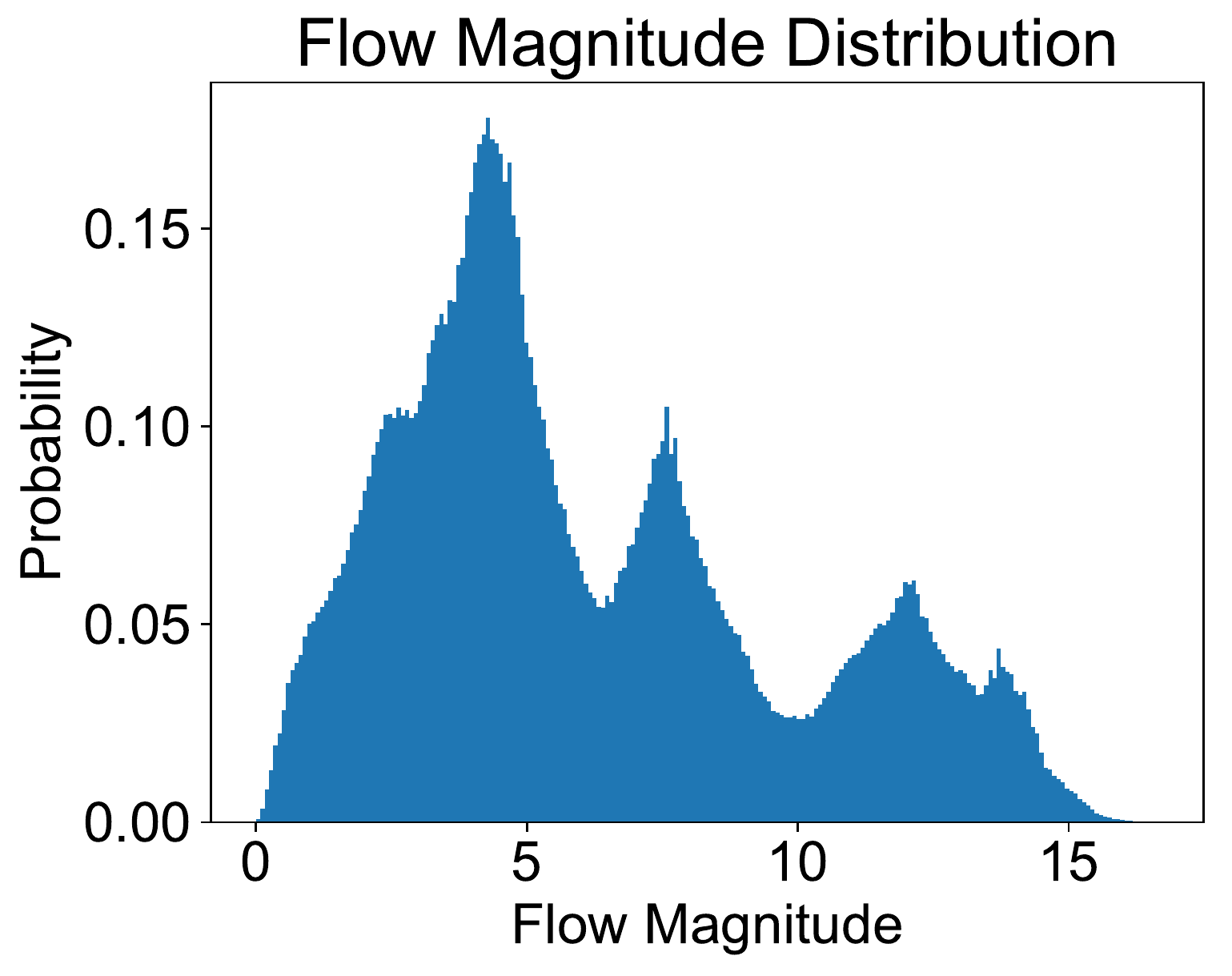}
      \caption{Magnitude distribution of the optical flow map (c).} 
  \end{subfigure}
\hfill    
  \begin{subfigure}[t]{0.235\textwidth}\footnotesize
  \centering
   \includegraphics[width=\linewidth]{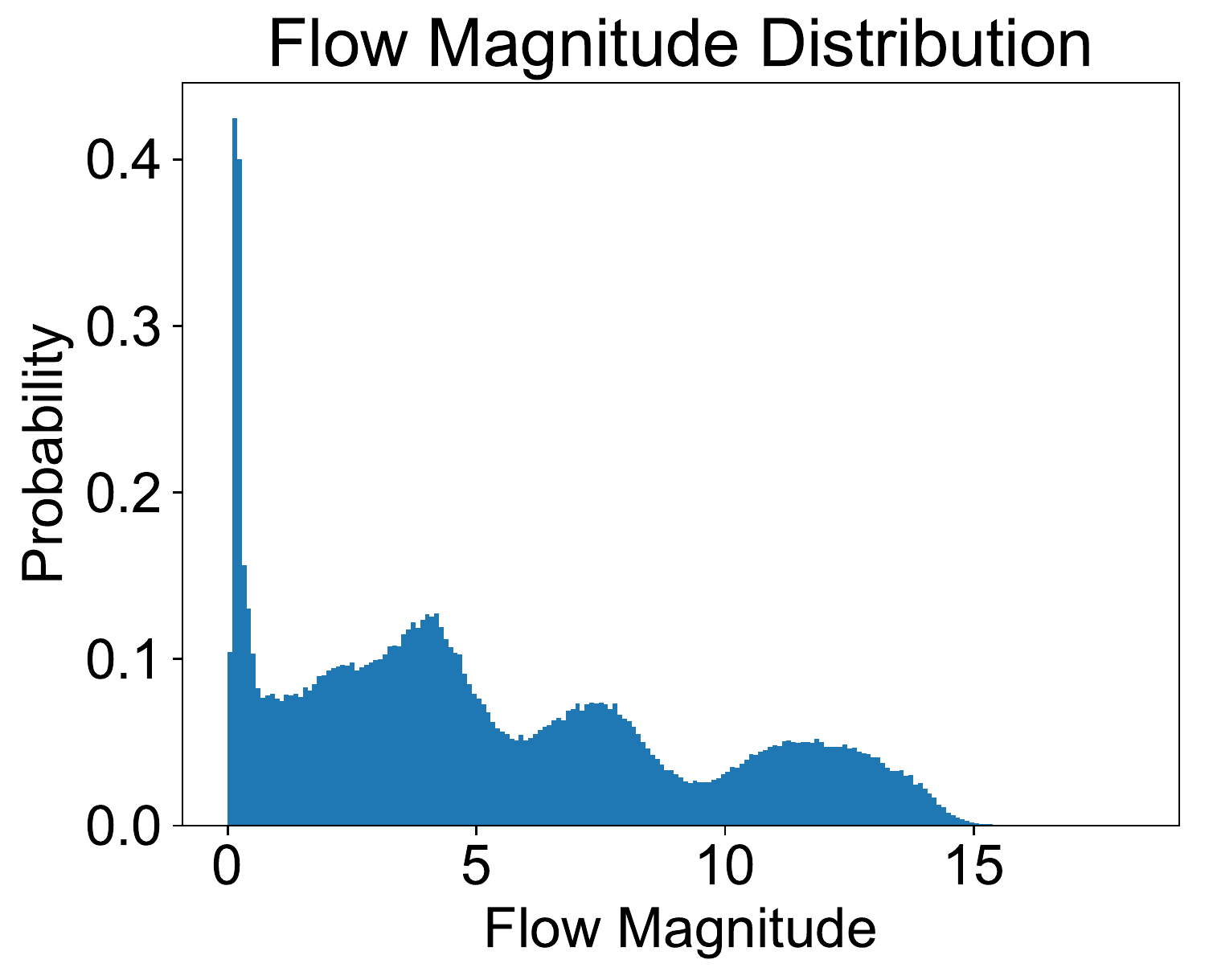}
      \caption{Magnitude distribution of the optical flow map (d).} 
  \end{subfigure}
  
  \caption{Flow visualize and statistic analysis. }
  
  \label{fig:flow_visual}
\end{figure}

In Fig. \ref{fig:flow_visual}, we provide further visual comparisons. 
Fig. \ref{fig:flow_visual} (a) and (b) represent the frame 5 and frame 6 from the Kimono sequence.
Fig. \ref{fig:flow_visual} (c) denotes the reconstructed optical flow map when the optical flow network is fixed during the training procedure. 
Fig. \ref{fig:flow_visual} (d) represents the reconstructed optical flow map after using the joint training strategy.
Fig. \ref{fig:flow_visual} (e) and (f) are the corresponding probability distributions of optical flow magnitudes.
It can be observed that the reconstructed optical flow by using our method contains more pixels with zero flow magnitude (e.g., in the area of human body). 
Although zero value is not the true optical flow value in these areas, our method can still generate accurate motion compensated results in the homogeneous region.
More importantly, the optical flow map with more zero magnitudes requires much less bits for encoding.
For example, it needs 0.045bpp for encoding the optical flow map in Fig. \ref{fig:flow_visual} (c) while it only needs 0.038bpp for encoding optical flow map in Fig. \ref{fig:flow_visual} (d). 

It should be mentioned that in the H.264 \cite{wiegand2003overview} or H.265 \cite{sullivan2012overview}, a lot of motion vectors are assigned to zero for achieving better compression performance.
Surprisingly, our proposed framework can learn a similar motion distribution without relying on any complex hand-crafted motion estimation strategy as in \cite{wiegand2003overview, sullivan2012overview}.

\begin{table}
    \centering
    \begin{tabular}{|cc|cc|cc|}
        \hline \multicolumn{2}{|c|}{Fix ME} & \multicolumn{2}{|c|}{W/O MVC} & \multicolumn{2}{|c|}{Ours}\\
        \hline Bpp & PSNR & Bpp & PSNR & Bpp & PSNR\\        
        \hline 0.044&27.33& 0.20& 24.32 & 0.029&  28.17\\
        \hline
    \end{tabular}
    \caption{The bit cost for encoding optical flow and the corresponding PSNR of warped frame from optical flow for different setting are provided.}
    \label{tab:jointMV}
\end{table}

\textbf{Motion Compensation.}
In this paper, the motion compensation network is utilized to refine the warped frame based on the estimated optical flow.
To evaluate the effectiveness of this module, we perform another experiment by removing the motion compensation network in the proposed system.
Experimental results of the alternative approach denoted by \emph{W/O MC} (see the \textcolor{green}{green curve} in Fig. \ref{fig:ablation}) show that the PSNR without the motion compensation network will drop by about 1.0 dB at the same bpp level. 

\textbf{Updating Strategy.}
As mentioned in Section \ref{sec:training}, we use an on-line buffer to store previously reconstructed frames $\hat{x}_{t-1}$ in the training stage when encoding the current frame $x_t$.
We also report the compression performance when the previous reconstructed frame $\hat{x}_{t-1}$ is directly replaced by the previous original frame $x_{t-1}$ in the training stage. 
This result of the alternative approach denoted by \textit{W/O update} (see the \textcolor{red}{red curve} ) is shown in Fig. \ref{fig:ablation}. 
It demonstrates that the buffering strategy can provide about 0.2dB gain at the same bpp level.

\textbf{MV Encoder and Decoder Network.}
In our proposed framework, we design a CNN model to compress the optical flow and encode the corresponding motion representations.
It is also feasible to directly quantize the raw optical flow values and encode it without using any CNN. 
We perform a new experiment by removing the MV encoder and decoder network. The experimental result in Fig. \ref{fig:ablation} shows that
the PSNR of the alternative approach denoted by \emph{W/O MVC} (see the \textcolor{magenta}{magenta curve} ) will drop by more than 2 dB after removing the motion compression network.
In addition, the bit cost for encoding the optical flow in this setting and the corresponding PSNR of the warped frame are also provided in Table \ref{tab:jointMV} (denoted by W/O MVC).
It is obvious that it requires much more bits (0.20Bpp) to directly encode raw optical flow values and the corresponding PSNR(24.43dB) is much worse than our proposed method(28.17dB). Therefore, compression of motion is crucial when optical flow is used for estimating motion.

\textbf{Motion Information.}
In Fig. \ref{fig:overview}(b), we also investigate the setting which only retains the residual encoder and decoder network. 
Treating each frame independently without using any motion estimation approach (see the \textcolor[rgb]{0.749,0.749,0.239}{yellow curve} denoted by \emph{W/O Motion Information}) leads to more than 2dB drop in PSNR when compared with our method.

\begin{figure}[!t]
\captionsetup[subfigure]{aboveskip=-0.8pt,belowskip=-1pt}
  \centering

  \begin{subfigure}[t]{0.235\textwidth}\footnotesize
  \centering
  \includegraphics[width=\linewidth]{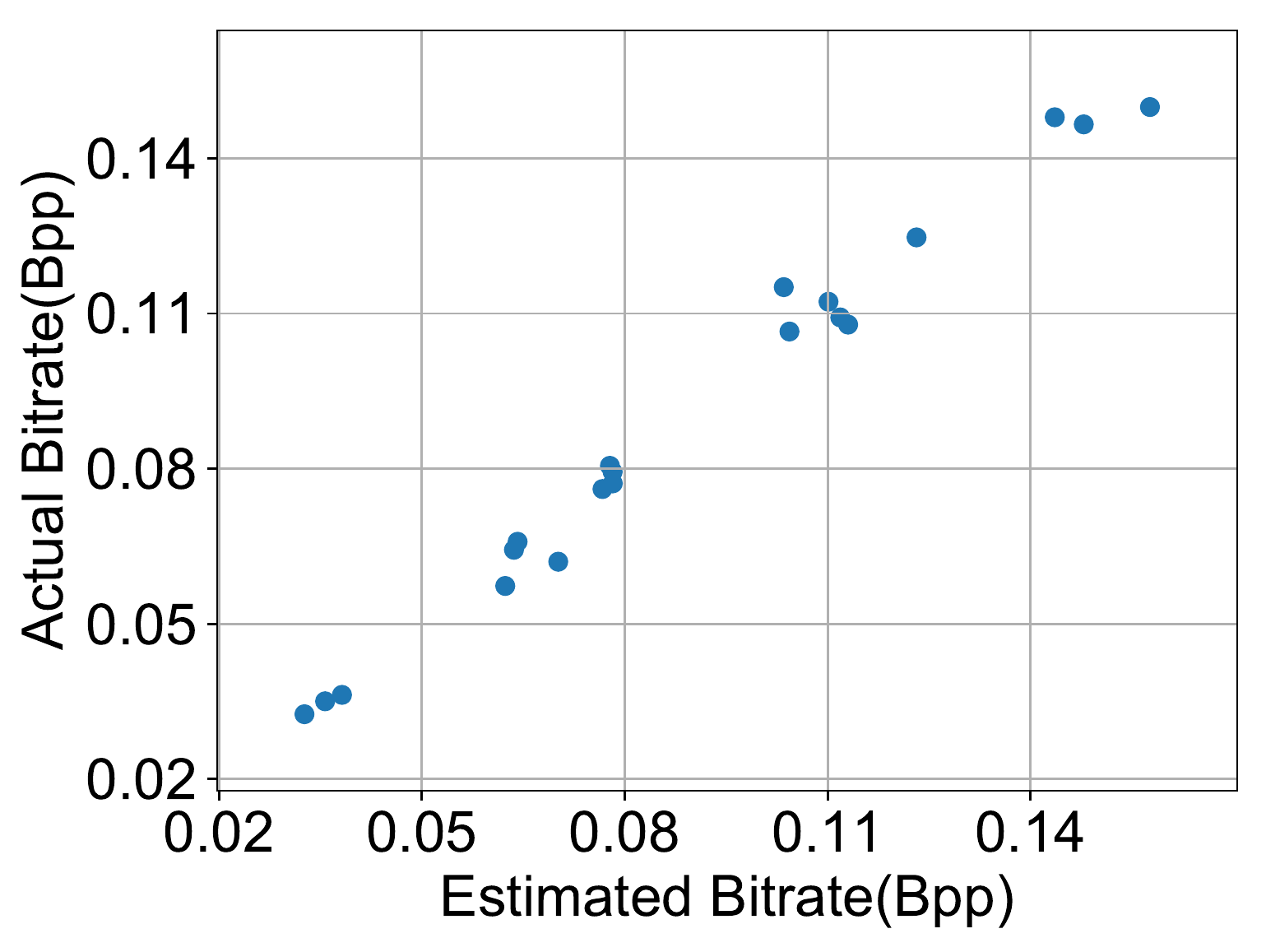}
      \caption{Actual and estimated bit rate.} 
  \end{subfigure}
\hfill    
  \begin{subfigure}[t]{0.235\textwidth}\footnotesize
  \centering
  \includegraphics[width=\linewidth]{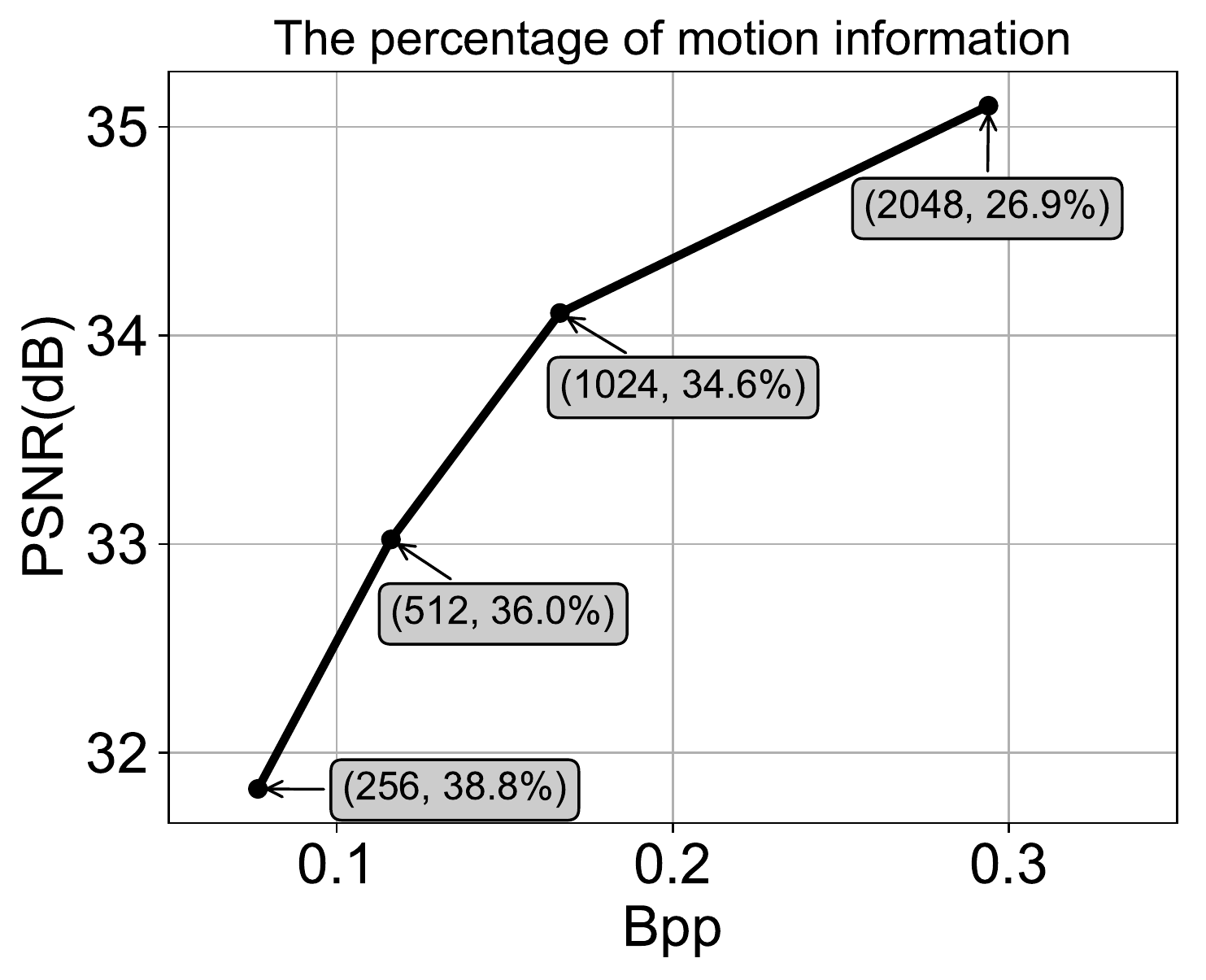}
      \caption{Motion information percentages.} 
  \end{subfigure}
  
  \caption{Bit rate analysis. }
  \label{fig:bitrate_analysis}
\end{figure}

\textbf{Running Time and Model Complexity.}
The total number of parameters of our proposed end-to-end video compression framework is about 11M. 
In order to test the speed of different codecs, we perform the experiments using the computer with Intel Xeon E5-2640 v4 CPU and single Titan 1080Ti GPU.
For videos with the resolution of 352x288, the encoding (\textit{resp.} decoding) speed of each iteration of Wu \textit{et al.}'s work \cite{Wu_2018_ECCV} is 29fps (\textit{resp.} 38fps), while the overall speed of ours is 24.5fps (\textit{resp.} 41fps).
The corresponding encoding speeds of H.264 and H.265 based on the official software JM \cite{JM} and HM \cite{HM} are 2.4fps and 0.35fps, respectively.
The encoding speed of the commercial software x264 \cite{x264} and x265 \cite{x265} are 250fps and 42fps, respectively.
Although the commercial codec x264 \cite{x264} and x265 \cite{x265} can provide much faster encoding speed than ours, they need a lot of code optimization. Correspondingly, recent deep model compression approaches are off-the-shelf for making the deep model much faster, which is beyond the scope of this paper. 

\textbf{Bit Rate Analysis.}
In this paper, we use a probability estimation network in \cite{balle2018variational} to estimate the bit rate for encoding motion information and residual information.
To verify the reliability, we compare the estimated bit rate and the actual bit rate by using arithmetic coding in Fig. \ref{fig:bitrate_analysis}(a).
It is obvious that the estimated bit rate is closed to the actual bit rate.
In addition, we further investigate on the components of bit rate.
In Fig. \ref{fig:bitrate_analysis}(b), we provide the $\lambda$ value and the percentage of motion information at each point. When $\lambda$ in our objective function $\lambda * D + R$ becomes larger, the whole Bpp also becomes larger while the corresponding percentage of motion information drops.

\section{Conclusion}
In this paper, we have proposed the fully end-to-end deep learning framework for video compression. 
Our framework inherits the advantages of both classic predictive coding scheme in the traditional video compression standards and the powerful non-linear representation ability from DNNs.
Experimental results show that our approach outperforms the widely used H.264 video compression standard and the recent learning based video compression system.
The work provides a promising framework for applying deep neural network for video compression. 
Based on the proposed framework, other new techniques for optical flow, image compression, bi-directional prediction and rate control can be readily plugged into this framework.

\noindent
\textbf{Acknowledgement}
This work was supported in part by National Natural Science Foundation of China (61771306)  Natural Science Foundation of Shanghai(18ZR1418100), Chinese National Key S\&T Special Program(2013ZX01033001-002-002), Shanghai Key Laboratory of Digital Media Processing and Transmissions(STCSM 18DZ2270700).

{\small
\bibliographystyle{ieee}
\bibliography{CVPR2019_FinalV1}
}

\clearpage
\newpage
\setcounter{equation}{0}
\setcounter{figure}{0}
\setcounter{table}{0}
\setcounter{page}{1}
\makeatletter

\begin{appendices}

\begin{table*}[ht]
\caption{BDBR and BD-PSNR(BD-MSSSIM) performances of H.265 and our DVC model when compared with H.264. }
\begin{center}
\begin{tabular}{|c|c|c|c|c|c|c|c|c|c|}
    \hline
    \multicolumn{2}{|c}{\multirow{3}{*}{Sequences}}&\multicolumn{4}{|c|}{H.265} & \multicolumn{4}{|c|}{Ours} \\
    \cline{3-10}
    \multicolumn{2}{|c}{}&\multicolumn{2}{|c|}{PSNR}&\multicolumn{2}{|c|}{MS-SSIM} &\multicolumn{2}{|c|}{PSNR}&\multicolumn{2}{|c|}{MS-SSIM}\\
    \cline{3-10}
    \multicolumn{2}{|c|}{}&\tabincell{c}{BDBR \\ (\%)} &\tabincell{c}{BD-\\PSNR(dB)} &\tabincell{c}{BDBR \\ (\%)} &\tabincell{c}{BD-MS\\SSIM(dB) }&\tabincell{c}{BDBR \\ (\%)} &\tabincell{c}{BD-\\PSNR(dB)} &\tabincell{c}{BDBR \\ (\%)} & \tabincell{c}{BD-MS\\SSIM(dB) }\\    
    \hline
    \multirow{5}{*}{ClassB} & BasketballDrive&-44.37&1.15&-39.80 & 0.87& -23.17 & 0.59 & -22.21 &0.51  \\
    &BQTerrace& -28.99&0.68 &-25.96 &0.50 &-25.12 & 0.54 &-19.52 &0.36\\
    &Cactus & -30.15 &0.68&-26.93& 0.47&-39.53& 0.94&-41.71&0.86 \\
    &Kimono & -38.81 &1.19&-35.31& 0.97& -40.70&1.23&-33.00&0.92\\
    &ParkScene &-16.35&0.45&-13.54&0.29&-25.20&0.77&-29.02&0.77\\
    \hline
    \multicolumn{2}{|c|}{Average} & \textbf{-31.73} &\textbf{0.83} &\textbf{-28.31} &\textbf{0.62} &\textbf{-30.75} &\textbf{0.81} &\textbf{-29.09} &\textbf{0.68}  \\
    
    \hline
    \multirow{4}{*}{ClassC} & BasketballDrill&-35.08&1.69&-34.04&1.41&-24.47&1.05&-27.18&1.18 \\
    &BQMall&-19.70&0.84&-17.57&0.60&26.13&-0.72&-18.85&0.67 \\
    &PartyScene &-13.41&0.60&-13.36&0.53&-9.14&0.29&-37.18&1.61\\
    &RaceHorses & -17.28&0.69&-17.01&0.57&-8.06&0.19&-29.24&1.05\\
    \hline
    \multicolumn{2}{|c|}{Average} & \textbf{-21.37}& \textbf{0.96}& \textbf{-20.50}& \textbf{0.78}& \textbf{-3.88}& \textbf{0.20}& \textbf{-28.11}& \textbf{1.13}  \\
    
    \hline
    \multirow{4}{*}{ClassD} & BlowingBubbles&-12.51&0.50&-10.28&0.35&-17.79&0.62&-35.44&1.53\\
    &BasketballPass&-19.26&0.99&-17.98&0.85&-0.39&-0.01&-20.53&1.01\\
    &BQSquare &-3.49&0.14&5.90&-0.19&-1.60&0.01&-23.67&0.84\\
    &RaceHorses & -14.77&0.68&-13.23&0.56&-18.95&0.72&-29.79&1.30\\
    \hline
    \multicolumn{2}{|c|}{Average} & \textbf{-12.51}&  \textbf{0.58}& \textbf{-8.89}& \textbf{0.39}&\textbf{-9.68}&\textbf{0.34}&\textbf{-27.36}&\textbf{1.17}\\
    
    \hline
    \multirow{3}{*}{ClassE} & Vidyo1&-37.12&1.11&-31.67&0.55&-36.05&1.20&-36.80&0.72\\
    &Vidyo3&-34.99&1.23&-29.48&0.65&-32.58&1.25&-40.09&1.02\\
    &Vidyo4 &-34.71&1.05&-27.41&0.61&-30.84&1.03&-24.84&0.66\\
    \hline
    \multicolumn{2}{|c|}{Average} & \textbf{-35.61}& \textbf{1.13}& \textbf{-29.52}&\textbf{0.61}&\textbf{-33.16}&\textbf{1.16}&\textbf{-33.91}&\textbf{0.80}\\

    \hline
    \multicolumn{2}{|c|}{Average Over All Sequences} & \textbf{-25.06}& \textbf{0.85}& \textbf{-21.73}& \textbf{0.60}& \textbf{-19.22}& \textbf{0.61}&\textbf{-29.32}&\textbf{0.94}\\    
    
    \hline
\end{tabular}
\end{center}
\label{tab:bdbr_hevc}
\end{table*}

\section{Experimental Results}
\subsection{BDBR and BD-PSNR(BD-MSSSIM) Results}

In the video compression task, BDBR and BD-PSNR (BD-MSSSIM) are widely used to evaluate the performance [1] of different video compression systems.
BDBR represents the average percentage of bit rate savings when compared with the baseline algorithm at the same PSNR (MS-SSIM).
BD-PSNR (BD-MSSSIM) represents the gain (dB) when compared with the baseline algorithm at the same bit rate.

In Table \ref{tab:bdbr_hevc}, we provide the BDBR and BD-PSNR(BD-MSSSIM) results of H.265 and the our proposed method DVC when compared with H.264.
When the distortion is measured by PSNR, our proposed model saves 19.22\% bit rate, while H.265 saves 25.06\% bit rate.
When measured by MS-SSIM, our proposed method can save more than 29\% bit rate, while H.265 only saves 21.73\%.
It clearly demonstrates that our proposed method outperforms H.264 in terms of PSNR and MS-SSIM.
Meanwhile, the performance of our method is comparable or even better than H.265 in terms of PSNR and MS-SSIM, respectively.

\subsection{Compression Performance on the HEVC Class C and Class D}

\begin{figure*}[htpb]
  \centering
  \begin{minipage}{0.4\textwidth}\footnotesize
    \centerline{\includegraphics[width=\linewidth]{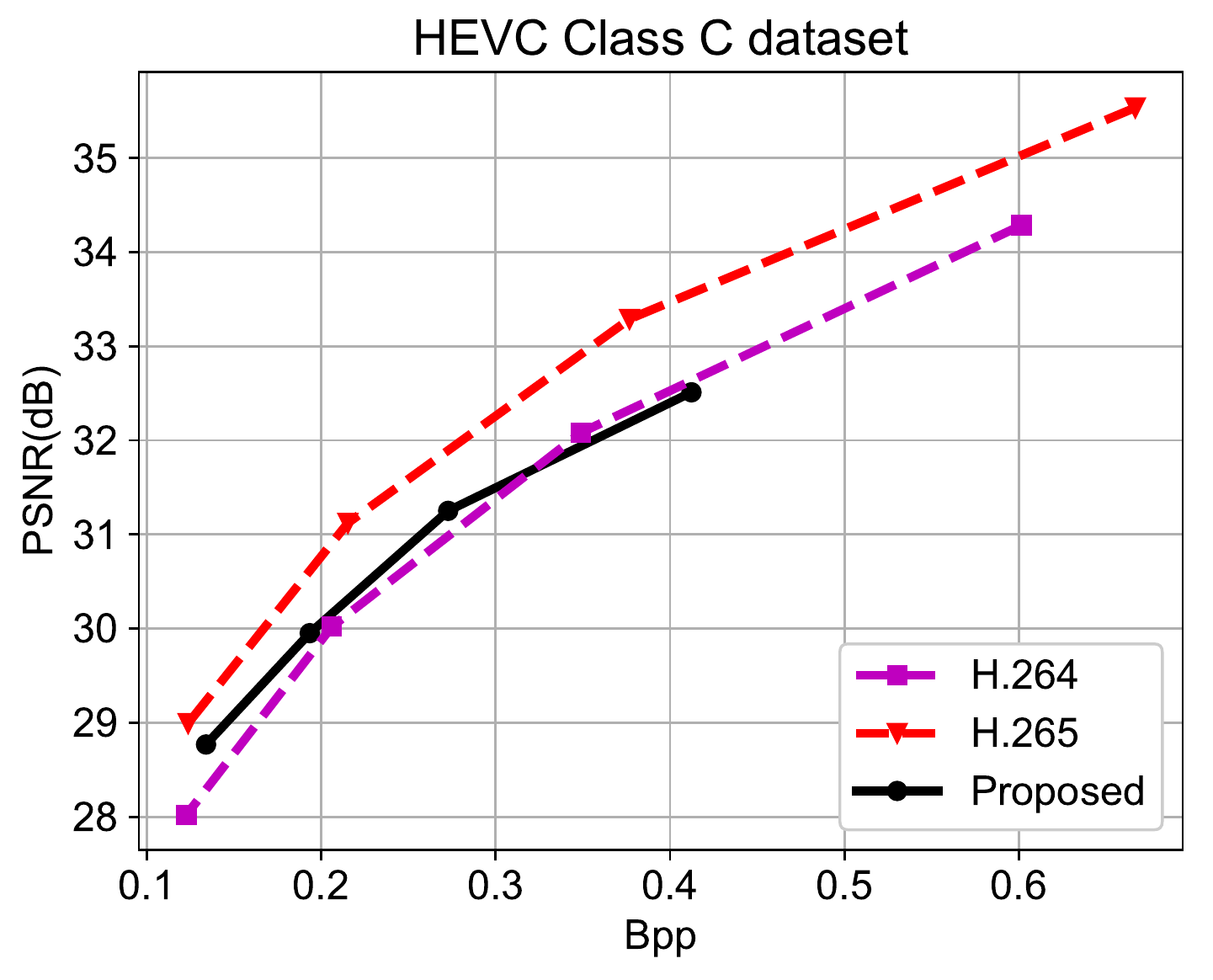}}
  \end{minipage}
\hspace{-0.2cm}
  \begin{minipage}{0.4\textwidth}\footnotesize
    \centerline{\includegraphics[width=\linewidth]{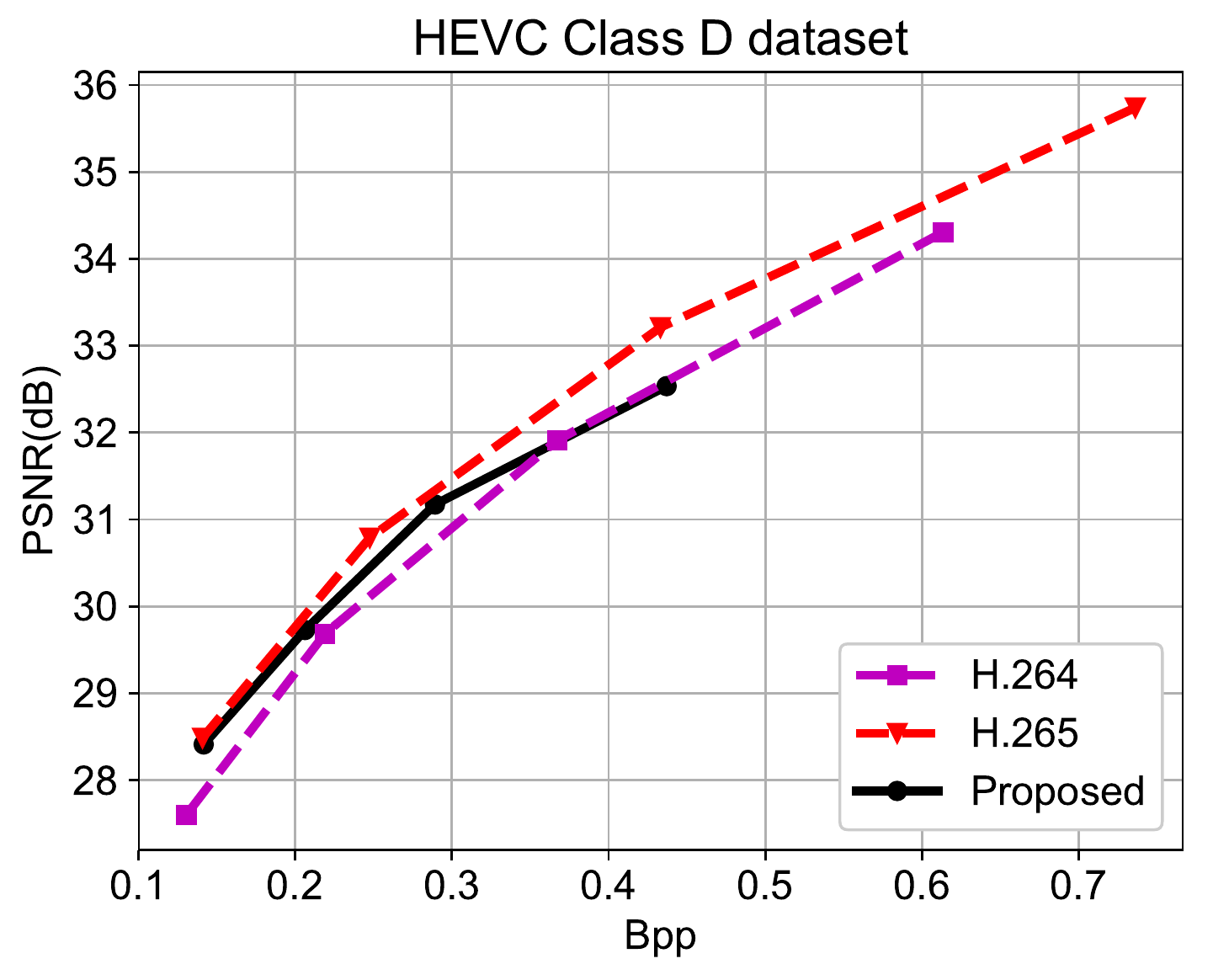}}
  \end{minipage}  
  
    \begin{minipage}{0.4\textwidth}\footnotesize
    \centerline{\includegraphics[width=\linewidth]{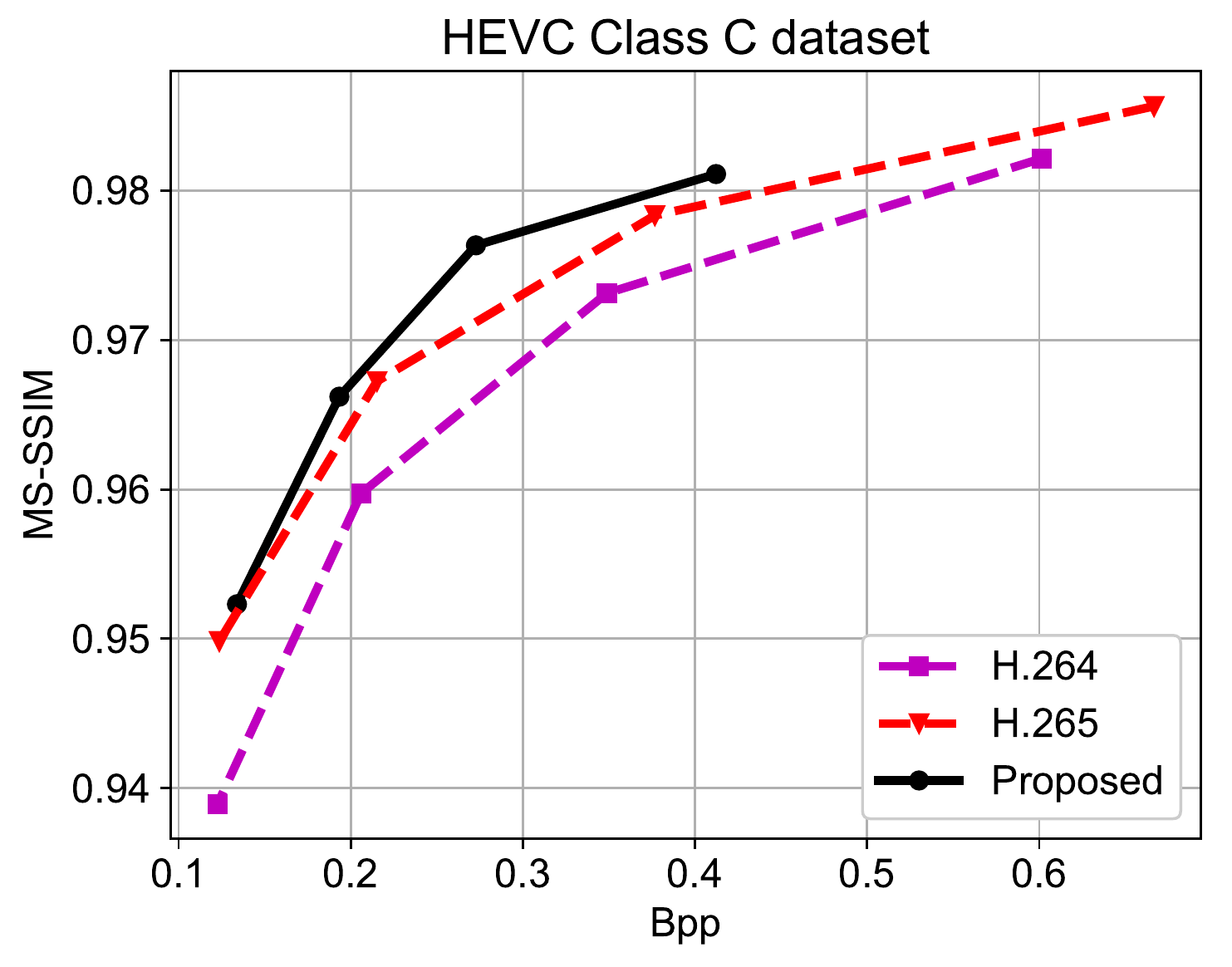}}
  \end{minipage}
\hspace{-0.2cm}
  \begin{minipage}{0.4\textwidth}\footnotesize
    \centerline{\includegraphics[width=\linewidth]{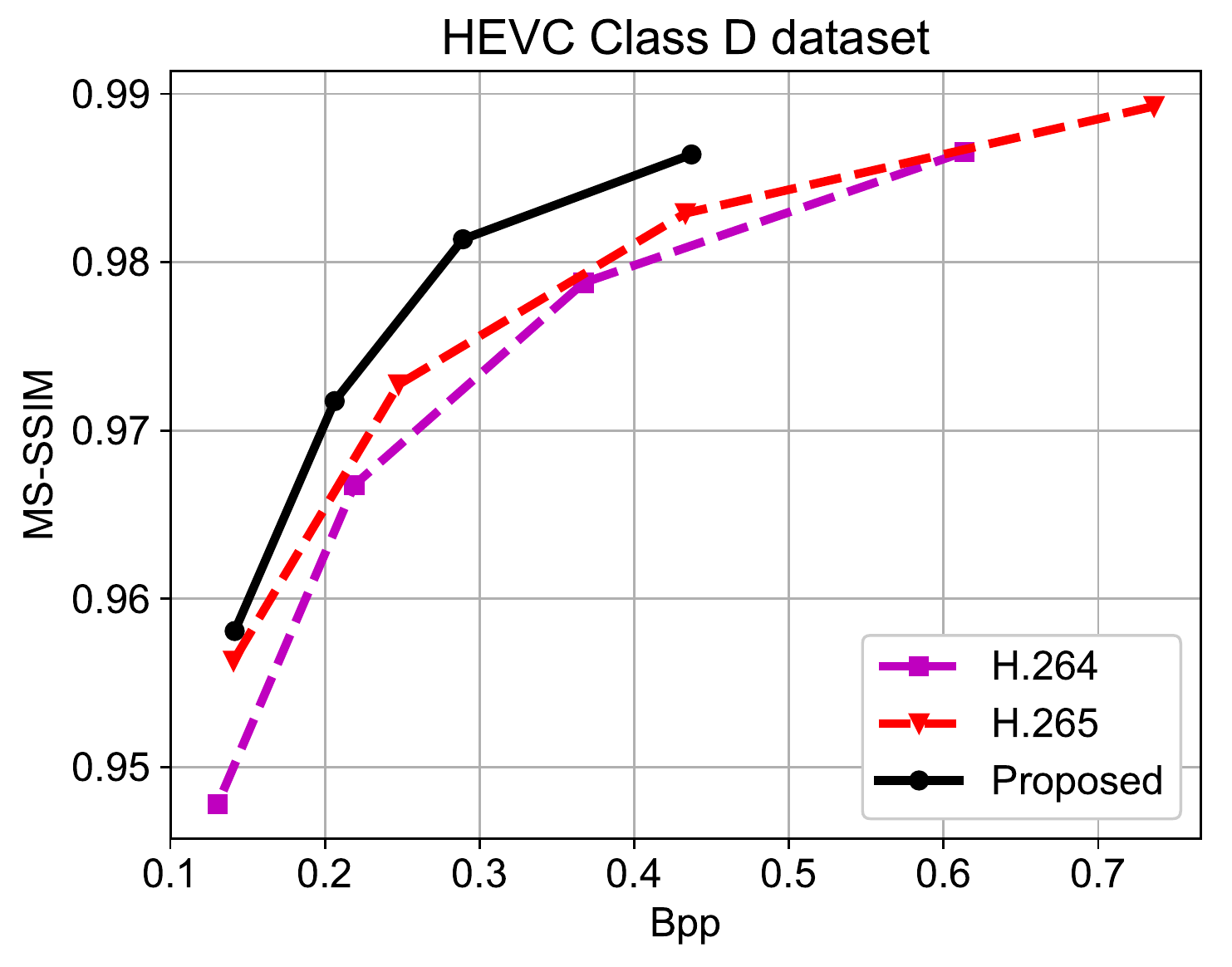}}
  \end{minipage} 
  
  \caption{Compassion of our method with the traditional video compression methods H.264 and H.265 on the HEVC Class C and Class D datasets.}
  \label{fig:performance}
\end{figure*}

In the submitted paper, we provide the performance on the HEVC Class B and Class E datasets.
In this section, we also compare our proposed method with the traditional video compression algorithms H.264 and H.265 on the HEVC Class C and Class D datasets in Figure \ref{fig:performance}.
Our proposed method still outperforms the H.264 algorithm in terms of PSNR and MS-SSIM.
According to Table \ref{tab:bdbr_hevc}, the newly proposed DVC model saves 3.88\% and 9.68\% bit rates when compared with H.264 on the HEVC Class C and Class D datasets, respectively.

\subsection{Result Using Kinetics Dataset as the Train Dataset}

\begin{figure}[htpb]
  \centering
  \begin{minipage}{0.4\textwidth}\footnotesize
    \centerline{\includegraphics[width=\linewidth]{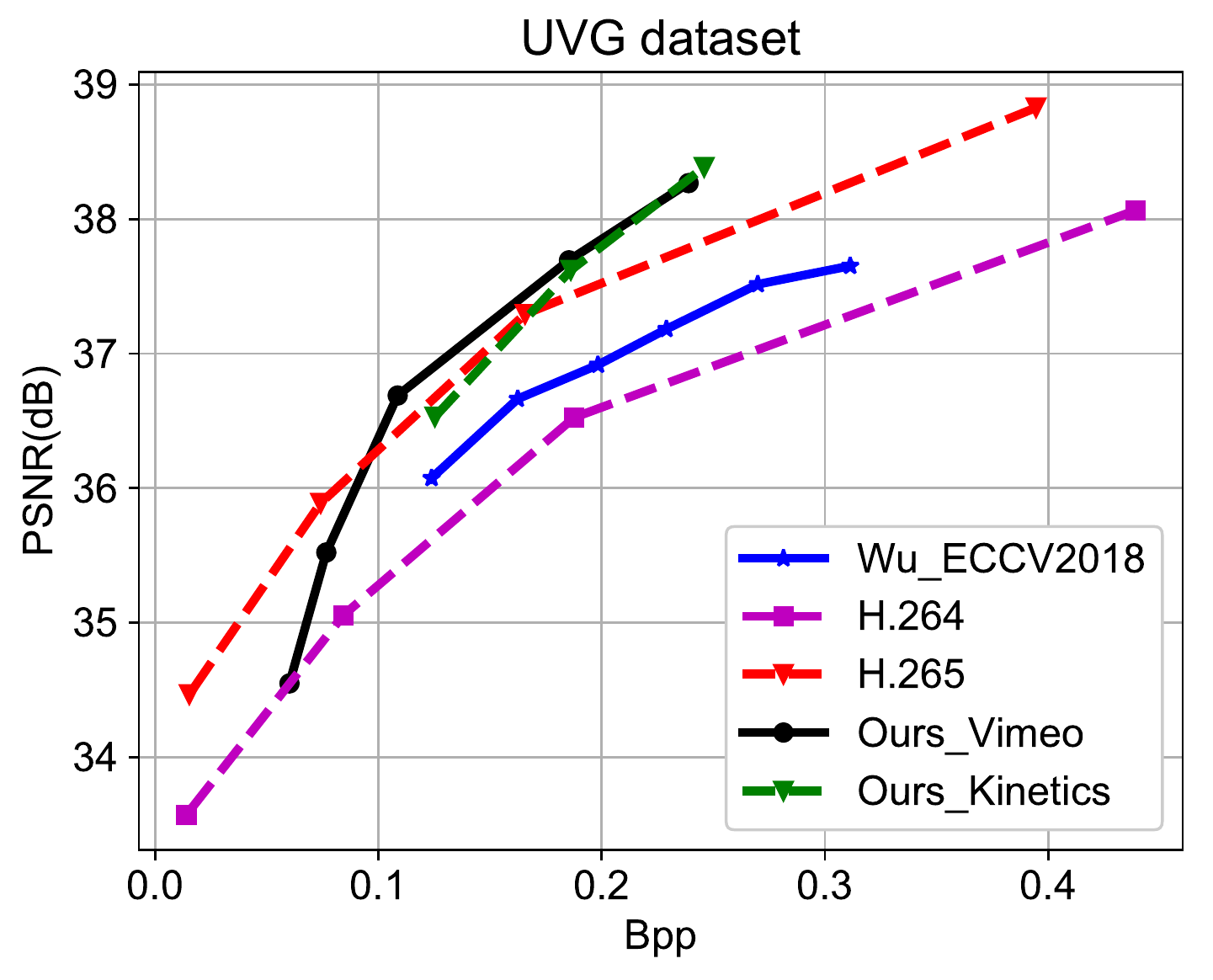}}
  \end{minipage}
  
  \caption{Compassion of our method with the traditional video compression methods H.264 and H.265 on the UVG dataset.}
  \label{fig:kinetics}
\end{figure}

Wu \textit{et al.} utilized the Kinetics datasets to train the video compression mdoel in [2].
In Figure \ref{fig:kinetics}, we also report the performance of our model when using the Kinetics dataset as the train dataset, which is denoted as \textit{Ours\_Kinetics} (see the \textcolor{green}{green curve}).
It is obvious that our method trained based on Kinetics dataset outperforms the H.264 and the baseline method [2].

\section{Experimental Settings}
\subsection{H.264 and H.265 Setting}
In order to generate the compressed videos from H.264 and H.265, we follow the setting in [2] (confirmed from the first author) and use the FFmpeg with the \textit{very fast} mode. 
Given the uncompressed video sequence \textit{Video.yuv} with the resolution of $W \times H$, the command line for generating H.264 compressed video is provided as follows,

\textit{
ffmpeg -y -pix\_fmt yuv420p -s WxH -r FR -i Video.yuv  -vframes N -c:v libx264 -preset veryfast -tune zerolatency -crf Q -g GOP-bf 2 -b\_strategy 0 -sc\_threshold 0  -loglevel debug output.mkv }

The command line for generating H.265 is provided as follows,

\textit{
ffmpeg  -pix\_fmt yuv420p -s WxH -r FR -i Video.yuv  -vframes N   -c:v libx265  -preset veryfast -tune zerolatency  -x265-params "crf=Q:keyint=GOP:verbose=1"  output.mkv  }

Among them, $FR, N, Q, GOP$ represents the frame rate, the number of encoded frames, quality, GOP size, respectively.
$N$ is set to 100 for the HEVC datasets. 
$Q$ is set as 15,19,23,27 in our settings.
$GOP$ is set as 10 for the HEVC dataset and 12 for the UVG dataset.

\subsection{PSNR, MS-SSIM and Bpp}
In our experiments, PSNR and MS-SSIM are evaluated on the RGB channel.
We average the PSNRs from all vdieo frames in each sequence to obtain the corresponding PSNR of this sequence.
Given the compressed frame with the resolution of $W \times H$ and the corresponding bit cost $R$, the Bpp is calculated as follows,
\begin{equation}
    Bpp  = R /W /H
\end{equation}

\section{Network Architecture of Our Motion Compensation Network}
\begin{figure}[!h]
  \centering
  \begin{minipage}{0.6\textwidth}\footnotesize
    \centerline{\includegraphics[width=\linewidth]{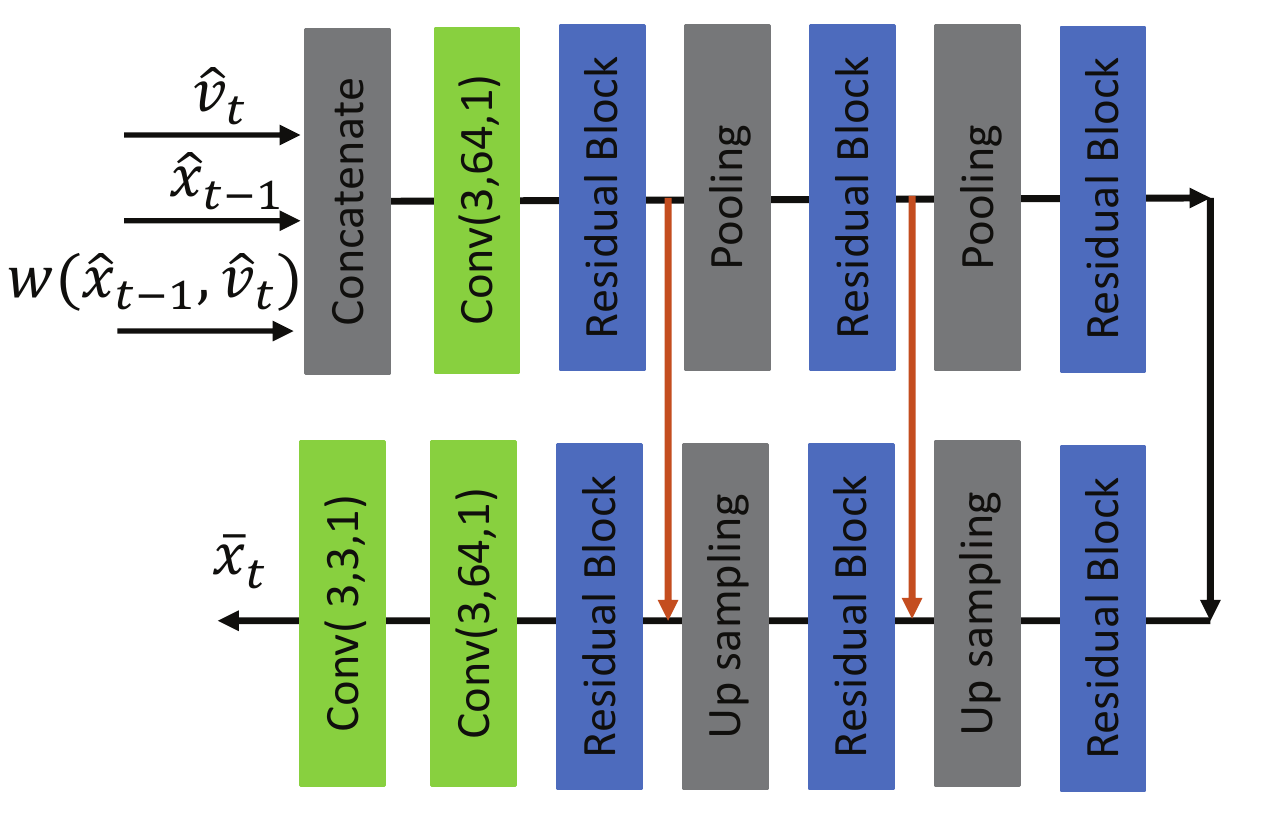}}
  \end{minipage}
  \caption{Network architecture of the motion compensation network.}
  \label{fig:mc}
\end{figure}

The motion compensation network is shown in Figure \ref{fig:mc}.
$w(\hat{x}_{t-1}, \hat{v}_t)$ represents the warped frame.
We use residual block with pre-activation structure.

\section{References}

[1] G. Bjontegaard. Calculation of average psnr differences between rd-curves. VCEG-M33, 2001.

[2] C.-Y. Wu, N. Singhal, and P. Krahenbuhl. Video compression through image interpolation. In ECCV, September 2018.

\end{appendices}

\end{document}